\documentclass[aps,prb,10pt,reprint,groupedaddress,showpacs,amssymb,amsfonts]{revtex4-1}
\usepackage{amsmath,amssymb,amscd,latexsym,epsfig,bm,subfigure}

\usepackage[usenames,dvipsnames]{color}
\usepackage[colorlinks,bookmarks=false,citecolor=blue,linkcolor=red,urlcolor=blue]{hyperref}
\usepackage{verbatim}

\def\beq{\begin{equation}}
\def\eeq{\end{equation}}
\def\be{\begin{equation}}
\def\ee{\end{equation}}

\def\ket#1{\vert #1 \rangle}

\newcommand{\zz}{\mathbb{Z}_2}
\newcommand{\z}{\mathbb{Z}}

\def\br{\boldsymbol{r}}

\def\cA{\mathcal{A}}
\def\cB{\mathcal{B}}
\def\cE{\mathcal{E}}

\begin{document}

\title{String flux mechanism for fractionalization in topologically ordered phases}

\author{Michael Hermele}
\affiliation{Department of Physics, 390 UCB, University of Colorado,
Boulder CO 80309, USA}
%\thanks{}
\date{\today}

\begin{abstract}
We construct a family of exactly solvable spin models that illustrate a novel mechanism for fractionalization in topologically ordered phases, dubbed the string flux mechanism.  The essential idea is that an anyon of a topological phase can be endowed with fractional quantum numbers when the string attached to it slides over a background pattern of flux in the ground state.  The string flux models that illustrate this mechanism are $\z_n$ quantum double models defined on specially constructed $d$-dimensional lattices, and possess $\z_n$ topological order for $d \geq 2$.  The models have a unitary, internal symmetry $G$, where $G$ is an arbitrary finite group. The simplest string flux model is a $\zz$ toric code defined on a bilayer square lattice, where $G = \zz$ is layer-exchange symmetry.  In general, by varying the pattern of $\z_n$ flux in the ground state, any desired fractionalization class [element of $H^2(G, \z_n)$] can be realized for the $\z_n$ charge excitations.  While the string flux models are not gauge theories, they map to $\z_n$ gauge theories in a certain limit, where they follow a novel \emph{magnetic} route for the emergence of low-energy gauge structure.  The models are analyzed by studying the action of $G$ symmetry on $\z_n$ charge excitations, and by gauging the $G$ symmetry.  The latter analysis confirms that distinct fractionalization classes give rise to distinct quantum phases, except that classes $[\omega], [\omega]^{-1} \in H^2(G, \z_n)$ give rise to the same phase.  We conclude with a discussion of open issues and future directions.
\end{abstract}

%\pacs{05.30.Pr, 75.10.Kt}

\maketitle

%\tableofcontents

\section{Introduction}
\label{sec:intro}

A remarkable property of topologically ordered\cite{wen89,wen90,wen13} states of matter is the existence of excitations with fractional quantum numbers.  In fractional quantum Hall (FQH) liquids, the Laughlin quasiparticles carry fractional electric charge,\cite{laughlin83}  which has been directly observed in experiment.\cite{depicciotto97,saminadayar97,Martin2004}  Beyond the FQH regime, many theories of topologically ordered quantum spin liquids possess $S = 1/2$ spinons.\cite{kalmeyer1987,wen89b,wen91,read91,sachdev92,balents99,senthil00,moessner01b}
  More generally, quantum number fractionalization is perhaps the simplest facet of \emph{symmetry enriched topological} (SET) phases,\cite{wen02,wen03,fwang06,kou08,kou09,yao10,huh11,gchen12,levin12,cho12,essin13,Mesaros2013,Hung2013,Lu2013,cwang13,Hung2013b,Xu2013,Hung2013c,Lu2013b,Huang2013,essin14,ygu14,ymlu14} which are states of matter characterized by non-trivial interplay between symmetry and topological order.

While quantum number fractionalization has been studied for a long time, only recently has a systematic understanding of it and other aspects of SET phases begun to emerge.  An equivalent, but perhaps more descriptive, term is \emph{symmetry fractionalization}, reflecting the fact that action of symmetry on the system fractionalizes into an action on individual anyon quasiparticles. Building on earlier works,\cite{wen02,kitaev06} A. M. Essin and I recently provided a classification of distinct types of symmetry fractionalization for Abelian topological orders\cite{essin13} (see also Ref.~\onlinecite{Mesaros2013}), which is an ingredient in the classification of SET phases.  For each type of anyon quasiparticle, the classification of Ref.~\onlinecite{essin13} assigns a \emph{fractionalization class} that describes the fractional action of symmetry and corresponding fractional quantum numbers.  The assignment of fractionalization classes to all anyon types specifies a \emph{symmetry class}, which is a robust property of a quantum phase, in the sense that it cannot change without either passing through a phase transition, or via explicit breaking of symmetry.  Mathematically, for an anyon that fuses with itself $n$ times to obtain a topologically trivial excitation, the distinct fractionalization classes are elements of the cohomology group $H^2(G, \z_n)$, where $G$ is the group of an internal, unitary symmetry.

Despite this and many other recent advances in the theory of topological phases, finding a topologically ordered phase beyond FQH liquids in a real system remains a major challenge.  To this end, it is important to discover microscopic mechanisms leading to topological phases, including SET phases in particular.  Ultimately, we would like to find mechanisms that operate in realistic models.  This is a challenging task, so it is valuable first to construct toy models that connect the general understanding provided by classifications of phases on the one hand, to concrete microscopic Hamiltonians on the other.

In this paper, we construct and study a family of exactly solvable Hamiltonians that illustrate a novel mechanism for symmetry fractionalization, which we dub the \emph{string flux} mechanism.  These string flux models directly encode the fractionalization classes of Ref.~\onlinecite{essin13} into a spin model Hamiltonian in arbitrary dimension $d$.  For $d \geq 2$, the models  exhibit $\z_n$ topological order, \emph{i.e.}, the topological order of the deconfined phase of $\z_n$ gauge theory.  
The  $\z_n$ charge excitations ($e$-particles) are endowed with fractional quantum numbers of an internal, unitary symmetry $G$, where $G$ is an arbitrary finite group.  Any desired fractionalization class in $H^2(G, \z_n)$ can be realized for $e$-particles, while the $\z_n$ flux excitations are always in the trivial fractionalization class.

The string flux mechanism builds on the string-net condensation mechanism for topological order,\cite{levin05} where the ground state of a topological phase is (in the simplest cases) visualized as a linear superposition of configurations of wildly fluctuating strings.  The ground states of our models are string-net condensates in which the wavefunction accumulates phase factors when strings slide over a static background configuration of $\z_n$ fluxes.  These phase factors are directly responsible for symmetry fractionalization. The string flux mechanism may potentially be useful in identifying new and more realistic models supporting topological order and fractionalization.

Beyond illustration of the string flux mechanism, the string flux models also illustrate a novel mechanism for the emergence of low-energy effective gauge theory.  In the usual mechanism for gauge theory to emerge at low energy, the Gauss' law constraint is imposed energetically, so we refer to this as the electric mechanism.  In a certain limit, the string flux models map to $\z_n$ gauge theory.  This limit does not impose Gauss' law, but instead involves energetic constraints on the flux excitations, and is thus a kind of \emph{magnetic} mechanism for emergence of gauge theory.  Further study of this mechanism may lead to new insights into how gauge structure can emerge at low energy in condensed matter systems.

While it is not our primary motivation, we would like to mention that our results provide an ``existence proof'' that all $H^2(G, \z_n)$ fractionalization classes are realized in local bosonic models with $\z_n$ topological order and finite, internal, unitary symmetry $G$.  This result can likely also be obtained using other classes of solvable models with topological order and fractionalization,\cite{chen14,Mesaros2013} which we discuss below.  The result could be obtained more easily -- but less rigorously -- using parton constructions; if one insists on a high degree of rigor, such constructions produce low-energy effective gauge theories, which are not themselves local bosonic models.

We now give an overview of our results, which also serves an outline of the paper, before closing this section with a brief discussion of related works.  To give a simple illustration of the string flux mechanism, we begin by presenting the simplest string flux model (Sec.~\ref{sec:simple}), which is a $\zz$ toric code\cite{kitaev03} on a specially constructed bilayer square lattice, where exchange of the two layers is a $G = \zz$ symmetry.  Depending on a parameter in the Hamiltonian that controls the pattern of $\zz$ flux in the ground state, the $e$-particles carry either integer or fractional $\zz$ charge.

In general, the string flux models are $\z_n$ quantum double models\cite{kitaev03} defined on specially constructed $d$-dimensional lattices, where the action of $G$ can be interpreted as an internal symmetry.  The Hamiltonian is a sum of commuting projectors.  Just like $\z_n$ gauge theories, to which they are intimately related, these models have $\z_n$ charge and flux excitations, and also admit the possibility of background patterns of charge and flux in the ground state.  Before the construction of the string flux models, Sec.~\ref{sec:background} presents background material on the class of models considered (so-called local bosonic models), $\z_n$ topological order, and the theory of symmetry fractionalization.  $\z_n$ quantum double models on a general graph are introduced in Sec.~\ref{sec:anygraph}.

To build the lattice on which the string flux models are defined, we start with $|G|$ copies of a $d$-dimensional hypercubic lattice; we borrow $d=2$ terminology and refer to these as layers.  Each layer is associated with a group element $g \in G$, and $G$ symmetry acts by permuting the layers according to group multiplication.  The layers are connected by introducing edges joining the $|G|$ vertices in each primitive cell of the hypercubic lattice.  The edges connecting layers form a \emph{Cayley graph} of $G$ in each primitive cell; this is a graph that represents group multiplication in $G$.\cite{lyndon77}

To specify the Hamiltonian, we choose a fractionalization class $[\omega] \in H^2(G, \z_n)$, which is associated with a $\z_n$-valued function of two group elements $\omega(g_1, g_2)$, called a factor set.  The factor set is encoded into the Hamiltonian in a natural way,  as a ground-state pattern of $\z_n$ fluxes passing through the cycles of each Cayley graph.  This is done first for a zero-dimensional quantum double model on a single Cayley graph in Sec.~\ref{sec:0d}, before the full construction of string flux models in Sec.~\ref{sec:gend}.  Section~\ref{sec:topo} shows that the string flux models in $d \geq 2$ have $\z_n$ topological order.  The $\z_n$ charge excitations ($e$-particles) reside at vertices of the lattice, and have symmetry fractionalization corresponding to the factor set $\omega(g_1, g_2)$.  This is established by exhibiting the operators that realize the fractional action of symmetry on a single $e$-particle, which are simply single spin operators of the quantum double model (Sec.~\ref{sec:sf2d}).

To gain further insight into the string flux models, we show that in a particular limit they map exactly to $\z_n$ gauge theories (Sec.~\ref{sec:gmapping}), which illustrates the magnetic mechanism for emergence of gauge theory discussed above.
The resulting gauge theory is a convenient starting point to study the string flux model ground states by 
  gauging the $G$ symmetry, which is a useful tool in the study of topological phases with unitary internal symmetry.\cite{levin12b}  Gauging $G$ symmetry produces a new gauge theory with (finite) gauge group $E$ (Sec.~\ref{sec:gauging}), where $E$ is the $\z_n$ central extension of $G$ associated with the fractionalization class $[\omega]$ (see Sec.~\ref{sec:gauging} for a definition).  This result can be anticipated following Ref.~\onlinecite{Hung2013}, and is consistent with results obtained  explicitly for $G = \z_2$ symmetry.\cite{Hung2013,Lu2013}  For the SET phases arising in the string flux models, we argue that isomorphism of central extensions (in an appropriate sense) corresponds to equivalence of SET phases.  This analysis confirms that string flux models with distinct fractionalization classes are in distinct SET phases, with the exception that fractionalization classes related by $\omega(g_1, g_2) \to [\omega(g_1, g_2)]^{-1}$ give rise to the same phase.  This can also be understood directly, without gauging symmetry, by noting that two such fractionalization classes are related by a relabeling of anyons (Sec.~\ref{sec:background}).
  
The paper concludes in Sec.~\ref{sec:discussion} with a discussion of open questions.  Several technical details are contained in appendices.

We close this section with a brief discussion of some related and prior work, including other families of exactly solvable models related to the string flux models.  A distinct route to SET phases with symmetry fractionalization proceeds by ``attaching'' $d=1$ symmetry protected topological (SPT) phases\cite{chen11a,turner11,fidkowski11,schuch11} to the fluctuating strings of a topologically ordered phase.\cite{yao10,chen13b,chen14}  This has the effect of attaching fractional quantum numbers to the ends of fluctuating strings, providing a mechanism for symmetry fractionalization that appears to be dual in some sense to the string flux mechanism discussed here.  In Ref.~\onlinecite{chen14}, these ideas were applied to construct certain Walker-Wang models\cite{walker12,keyserlingk13} for $d=3$ SPT phases.  To my knowledge, the analogous construction has not explicitly appeared in the literature for models with topological order (\emph{e.g.} quantum double models), but this is an immediate application of Ref.~\onlinecite{chen14}.  Applying this construction to $\z_n$ quantum double models seems likely to realize any desired fractionalization class for $e$-particles.  It will be interesting in future work to study the relationship of these models to string flux models, as well as these two different mechanisms for symmetry fractionalization.

A rather different family of models has been studied by Mesaros and Ran, who, building on the cohomology classification of $d=2$ SPT phases,\cite{chen13} constructed models with unitary internal symmetry $G_s$ and ``gauge group'' $G_g$, where each model encodes an element of $H^3(G_s \times G_g, {\rm U}(1))$.\cite{Mesaros2013}  Focusing on $G_s$ and $G_g$ Abelian, Mesaros and Ran studied symmetry fractionalization in these models, finding, in contrast to the string flux models constructed here, that the gauge flux excitations can have a non-trivial fractionalization class, while the gauge charge excitations are always trivial.  The Mesaros-Ran models are richer than string flux models, realizing phenomena beyond symmetry fractionalization.  However, the connection between the construction of the model and the type of symmetry fractionalization present is not transparent.  A closely related family of models was studied by Hung and Wen,\cite{Hung2013} who started with a SPT phase with $E$ symmetry, for $E$ a $G_g$ extension of $G_s$.  They gauged the $G_g$ subgroup of $E$ to obtain a SET phase with $G_s$ symmetry.  Studying the special case $G_s = G_g = \zz$, they found that both $\zz$ charge and flux excitations can have non-trivial fractionalization class in their models.

Finally, we mention an important precursor to this work, namely the projective symmetry group (PSG) approach to parton mean field theories of quantum spin liquids.\cite{wen02,wen03}  As discussed in Ref.~\onlinecite{essin13}, PSG is the mean-field analog of fractionalization class.  Indeed, the notion of fractionalization class can be viewed as an extension of PSG beyond mean-field theory.  Many parton mean-field states are characterized by patterns of background flux felt by the partons, that give rise to non-trivial PSGs.\cite{wen02}  This is a parton theory instance of the string flux mechanism; the strings are electric field lines of the gauge theory, which are hidden in mean-field theory because the conjugate magnetic field degrees of freedom are taken to be non-fluctuating.  Of particular importance, these ideas were developed further in Ref.~\onlinecite{wen03}, where, focusing on lattice translation symmetry, an explicit connection between PSG and string-net condensation was made.

\section{Simple example: string flux model with $\zz$ symmetry}
\label{sec:simple}

Before proceeding to the construction of the full family of models in Sec.~\ref{sec:gend}, in this section we introduce the simplest string flux model and study some of its properties.  The discussion of this section is informal; the results asserted follow as a special case of the more detailed and careful discussion given in the remainder of the paper.
The model is a version of Kitaev's toric code model that depends on a parameter $K = \pm 1$ and has $G = \zz$ symmetry, appearing in the general construction of Sec.~\ref{sec:gend} for $n=2$ and $G = \zz$.  For both values of $K$, the model has $\zz$ topological order.  

For $K = -1$ ($K = 1$), the $e$-particles carry fractional (integer) charge under the global $\zz$ symmetry.  Letting $\zz = \{ 1, a \}$, we let $U^e_a$ be a unitary operator giving the action of the non-trivial element $a \in \zz$ on a single $e$-particle.  Fractional $\zz$ charge means $(U^e_a)^2 = -1$, while integer $\zz$ charge means $(U^e_a)^2 = 1$.  This can be made more intuitive by thinking about a situation with ${\rm U}(1)$ symmetry, where $e$-particles could have either integer or half-odd integer charge.  If the ${\rm U}(1)$ is then broken down to $\zz$, integer ${\rm U}(1)$ charge becomes integer $\zz$ charge, while half-odd integer ${\rm U}(1)$ charge becomes fractional $\zz$ charge.

\begin{figure}[t]
\includegraphics[width=0.9\columnwidth]{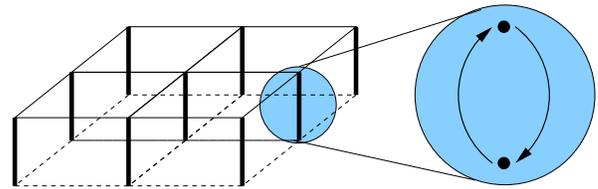}
\caption{(Color online.) Bilayer square lattice on which the simplest model with $G = \zz$ symmetry and $\zz$ topological order is defined.  The degrees of freedom are spin-1/2 spins residing on links.  Each pair of vertically adjacent sites is connected by two distinct links, labeled with up and down arrows.}
\label{fig:simple-model}
\end{figure}

The model is defined on the lattice shown in Fig.~\ref{fig:simple-model}.  This is a bilayer square lattice, where each pair of vertically adjacent sites is connected by not one but two links.  These links are labeled with ``up'' and ``down'' arrows in Fig.~\ref{fig:simple-model}.  On each link, we place a spin-1/2 spin, and the model is built from Pauli matrices $\sigma^\mu_\ell$ ($\mu = x,y,z$) acting on the spin at link $\ell$.

\begin{figure}
\includegraphics[width=0.5\columnwidth]{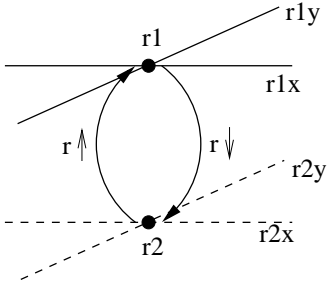}
\caption{Detail of the bilayer square lattice, illustrating the labeling of sites and links.}
\label{fig:sm-labels}
\end{figure}

We now introduce some notation to label the links and vertices of the lattice (see Fig.~\ref{fig:sm-labels}).  Each square lattice primitive cell is labeled by $\br = n_x \hat{e}_x + n_y \hat{e}_y$, for integers $n_x, n_y$, with $\hat{e}_x$ and $\hat{e}_y$ unit vectors in the $x$ and $y$ directions, respectively. The upper (lower) vertices are labeled by $\br 1$ ($\br  2$).  Links within each square layer are labeled by $\br 1 x, \br 1 y, \br 2 x, \br 2 y$.  Finally, links connecting the two layers are labeled by $\br \uparrow$ and $\br \downarrow$.

The $\zz$ symmetry acts by exchanging the two layers, and also exchanging $\br \uparrow \leftrightarrow \br \downarrow$.  In terms of operators, $U_a^2 = 1$ and
\begin{eqnarray}
U_a \sigma^{\mu}_{\br 1 \alpha} U^{-1}_a &=& \sigma^{\mu}_{\br 2 \alpha} \text{,} \\
U_a \sigma^{\mu}_{\br \uparrow} U^{-1}_a &=& \sigma^{\mu}_{\br \downarrow} \text{,}
\end{eqnarray}
where $\alpha = x,y$.  While it is convenient to visualize the symmetry spatially, it does not give rise to a translation or other rigid motion in two-dimensional space, and is properly considered an internal symmetry.

The Hamiltonian is simply the toric code model placed on the lattice described above, and is thus built from vertex operators that can be thought of as measuring a $\zz^g$ ``gauge'' charge at each vertex, and plaquette operators measuring  $\zz^g$ ``gauge'' flux through certain elementary cycles of the lattice.  It is important to keep in mind that $\zz^g$ is distinct from the symmetry group $G = \zz$.
  The vertex operator residing at $\br i$ ($i = 1,2$) is defined by
\begin{equation}
A_{\br i} = \prod_{\ell \sim \br i} \sigma^x_{\ell} \text{,}
\end{equation}
where the product is over the six links touching the vertex $\br i$.
There are three types (I, II, III) of plaquette operators. Type I plaquettes consist of the two links $\br \uparrow$, $\br \downarrow$ in each primitive cell $\br$:
\begin{equation}
B^{{\rm I}}_{\br} = \sigma^z_{\br \uparrow} \sigma^z_{\br \downarrow} \text{.}
\end{equation}
Type II plaquettes are associated with nearest-neighbor pairs of square lattice sites, and connect the top and bottom layers,
\begin{eqnarray}
B^{{\rm II}} _{\br \alpha \uparrow } &=& \sigma^z_{\br \uparrow} \sigma^z_{\br 1 \alpha} \sigma^z_{\br 2 \alpha} \sigma^z_{\br + \hat{e}_\alpha, \uparrow} \\
B^{{\rm II}}_{\br \alpha \downarrow} &=& \sigma^z_{\br \downarrow} \sigma^z_{\br 1 \alpha} \sigma^z_{\br 2 \alpha} \sigma^z_{\br + \hat{e}_\alpha, \downarrow}  \text{.}
\end{eqnarray}
Finally, type III plaquette operators are defined on the square faces in the top and bottom layers,
\begin{equation}
B^{{\rm III}}_{\br i} = \sigma^z_{\br i x} \sigma^z_{\br+\hat{e}_x, i y} \sigma^z_{\br + \hat{e}_y, i x} \sigma^z_{\br i y} \text{,} \quad i = 1,2 \text{.}
\end{equation}
The vertex and plaquette operators thus defined form a mutually commuting set of observables.  It is obvious that vertex (plaquette) operators commute with other vertex (plaquette) operators.  It is also true that $[A, B] = 0$ for any vertex operator $A$ and any plaquette operator $B$, since these operators share an even number of links.

The Hamiltonian is defined to be
\begin{eqnarray}
H &=& - \sum_{\br} \sum_{i = 1,2} A_{\br i} - \sum_{\br} \sum_{i = 1,2} B^{{\rm III}}_{\br i}  \nonumber \\ &-& \sum_{\br} \sum_{\alpha = x,y} [ B^{{\rm II}}_{\br \alpha \uparrow} + B^{{\rm II}}_{\br \alpha \downarrow} ] - K \sum_{\br} B^{{\rm I}}_{\br} \text{,}
\end{eqnarray}
where $K = \pm 1$.  $e$-particle excitations reside at vertices $\br i$ for which $A_{\br i} = -1$.

\begin{figure}
\includegraphics[width=\columnwidth]{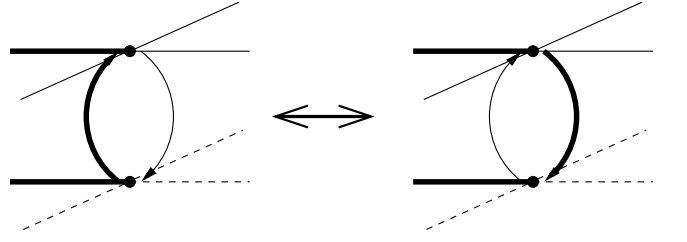}
\caption{Two string configurations that differ by the string (thick links) sliding over a type I plaquette.  The coefficients of these configurations in the ground state wavefunctions differ corresponding by a phase factor $K = \pm 1$.   The string feels a $\zz^g$ flux through the type I plaquette when $K = -1$.}
\label{fig:stringminus}
\end{figure}

We say that links with $\sigma^x = -1$ are occupied by a string, while links with $\sigma^x = 1$ have no string.  Ground states of $H$ are equal amplitude superpositions of all closed string configurations.  For $K = 1$, all string configurations have positive coefficient, while the coefficients alternate in sign for $K = -1$.  In particular, two string configurations that differ by a string sliding over a type I plaquette have coefficients differing by a minus sign, as shown in Fig.~\ref{fig:stringminus}.  Strings thus feel a pattern of non-trivial $\zz^g$ flux for $K = -1$.

These ground state phase factors are directly responsible for the fractional $\zz$ charge.  To see this, we let $U^e_a(\br i)$ be the operator giving the action of $a \in \zz$ on a single $e$-particle at site $\br i$.  We assert that
\begin{eqnarray}
U^e_a(\br 1) &=& \sigma^z_{\br \downarrow} \\
U^e_a(\br 2) &=& \sigma^z_{\br \uparrow} \text{.}
\end{eqnarray}
Then, for instance, for an $e$-particle at $\br 1$,
\begin{equation}
(U^e_a)^2 = U^e_a(\br 2) U^e_a(\br 1) = \sigma^z_{\br \uparrow} \sigma^z_{\br \downarrow} = B^{{\rm I}}_{\br} = K \text{,}
\end{equation}
where the last equality holds acting on any state for which the only excitations are $e$-particles (\emph{i.e.} no flux excitations).  Therefore, $K = -1$ corresponds to fractional $\zz$ charge, while $K = 1$ $e$-particles carry integer $\zz$ charge.  These are the two possible fractionalization classes given by $H^2(\zz, \zz^g) = \zz$, so both classes occur in the model upon tuning of $K$.

The distinction between $K = 1$ and $K = -1$ ground states can be further understood by gauging the $\zz$ symmetry, following the analysis of Secs.~\ref{sec:gmapping} and~\ref{sec:gauging}.  Upon gauging $\zz$, one obtains a $\zz \times \zz$ gauge theory for $K = 1$, and a $\z_4$ gauge theory for $K = -1$.  This clearly shows that the $K = 1$ and $K = -1$ ground states belong to different SET phases.

\section{Background: Local bosonic models, $\z_n$ topological order, and symmetry fractionalization}
\label{sec:background}

We begin by reviewing some background material that will be important in what follows.  We describe the class of systems to which we confine our attention (local bosonic models with an energy gap), and and briefly review $\z_n$ topological order in $d \geq 2$.  The focus is primarily on $d=2$, but we also discuss the higher-dimensional case.
We then describe those aspects of the theory of symmetry fractionalization that will be important below, following Ref.~\onlinecite{essin13}.

Throughout the paper, we confine our attention to local bosonic models, including for example lattice models of bosons or spins.  By definition, the Hilbert space in such systems is a tensor product of local Hilbert spaces, where each local Hilbert space describes degrees of freedom in some local region (\emph{e.g.} within a primitive cell of a crystal lattice).  
The interactions are local, meaning that degrees of freedom beyond some fixed spatial range are not coupled in the Hamiltonian.  In addition, we will be interested exclusively in systems with an energy gap to local excitations.  For simplicity, we consider systems with periodic boundary conditions unless stated otherwise.

$\z_n$ topological order in $d=2$ is characterized by the fusion and braiding properties of its anyon quasiparticle excitations.  Any localized excitation can be assigned one of $n^2$ particle types, which form a $\z_n \times \z_n$ fusion group generated by $e$ and $m$, both of which obey bosonic self-statistics.  The statistics is non-trivial because $e$ and $m$ obey $\theta = 2\pi / n$ mutual statistics.  The designation of $e$ and $m$ among the $n^2$ particle types is somewhat arbitrary, as there are non-trivial relabelings that preserve the fusion and braiding properties.  Two of these relabelings are
\begin{equation}
e \leftrightarrow m \text{,}
\end{equation}
and
\begin{eqnarray}
e &\leftrightarrow& e^{n-1}  \label{eqn:e-relabel} \\
m &\leftrightarrow& m^{n-1} \label{eqn:m-relabel} \text{.}
\end{eqnarray}
In fact, these generate all possible relabelings.  The latter relabeling will play an important role in some of our discussion.

The name $\z_n$ topological order arises from the fact that the deconfined phase of $\z_n$ lattice gauge theory in $d=2$ gives a realization of the properties described above.\footnote{More precisely, when we refer to the deconfined phase of $\z_n$ gauge theory, we mean that we start with a topologically trivial paramagnet with $\z_n$ global symmetry, which is then coupled to a dynamical but weakly fluctuating $\z_n$ gauge field.  We want to exclude the possibility that the $\z_n$ paramagnet is in a nontrivial SPT phase, which in general leads to a different topological order upon gauging the $\z_n$ symmetry.\cite{levin12b}  Put another way, this amounts to assuming that the (bosonic) matter sector of a $\z_n$ gauge theory is topologically trivial.}   It is sometimes useful to use gauge theory language, referring to $e$ particles as $\z_n$ charges, and $m$-particles as $\z_n$ fluxes, since this is how these excitations arise in the gauge theory.  In dimensions $d \geq 2$, we can define $\z_n$ topological order via the properties of the deconfined phase of $\z_n$ gauge theory.\cite{Note1}  Comparing to $d=2$, there are still point-like $\z_n$ charge ($e$-particle) excitations, and the $m$-particles become $(d-2)$-dimensional flux excitations (\emph{e.g.} $\z_n$ flux lines in $d=3$).  It is thus well-defined to bring an $e$-particle around a flux excitation, and this results in a $\theta = 2\pi / n$ statistical phase factor.  In gauge theory language, the relabeling of anyons given in Eqs.~(\ref{eqn:e-relabel}, \ref{eqn:m-relabel}) corresponds to the non-trivial automorphism of the $\z_n$ gauge group (where $a \to a^{-1}$ for $a \in \z_n$).  Physically, this corresponds to taking the inverse of all $\z_n$ charges and fluxes, and this transformation is defined in all $d \geq 2$.

Another simple realization of $\z_n$ topological order is in the exactly solvable $\z_n$ quantum double model\cite{kitaev03} on the $d$-dimensional hypercubic lattice ($d \geq 2$).  This model is intimately related to $\z_n$ gauge theory but, unlike the gauge theory, is a local bosonic model.  $\z_n$ quantum double models are introduced in detail in Sec.~\ref{sec:anygraph}, as the models constructed in this paper are of this type, defined on special lattices where $G$ acts as an internal symmetry.

In this paper, we will be almost entirely concerned with \emph{only} $e$-particles.  The property of greatest importance will be the fusion rule $e^n = 1$, which holds for all $d \geq 2$.  This fusion rule implies that single isolated $e$-particles cannot be locally created, but a group of $n$ $e$-particles can be created locally and then separated.  Physical excited states must therefore contain only multiples of $n$ $e$-particles.

Our purpose is to consider the interplay of $\z_n$ topological order with symmetry.  We restrict attention to unitary internal symmetry, which means that symmetry operations are represented in Hilbert space as a tensor product of unitary operators acting on the local Hilbert spaces.\footnote{This type of symmetry is often referred to as ``on-site symmetry,'' but we prefer the term internal symmetry.}
Physical examples of internal symmetry include spin rotation symmetry, ${\rm U}(1)$ charge symmetry, and time reversal.  We will make the further restriction that the symmetry group $G$ is finite.  

The action of  $g \in G$ on Hilbert space is represented by the unitary operator $U_g$. For any operator ${\cal O}$, we make the crucial assertion that
\begin{equation}
U_{g_1} U_{g_2} {\cal O} U^{-1}_{g_2} U^{-1}_{g_1} = U_{g_1 g_2} {\cal O} U^{-1}_{g_1 g_2} \text{.} \label{eqn:weak-linear-action}
\end{equation}
Na\"{\i}vely, we might imagine that this equation only holds up to a phase factor, which would be a projective action of $G$ on local operators.  However, in any physically reasonable model, symmetry must act linearly (\emph{i.e.} not projectively) on operators, as expressed in Eq.~(\ref{eqn:weak-linear-action}).  This is true because the property~(\ref{eqn:weak-linear-action}) holds for all physical (electrically neutral) bosonic degrees of freedom that can be microscopic constituents of a condensed matter system, such as electron spins or bosonic atoms.  The models we study here also turn out to have the stronger property that
\begin{equation}
U_{g_1} U_{g_2} = U_{g_1 g_2} \text{.} \label{eqn:strong-linear-action}
\end{equation}

Following Ref.~\onlinecite{essin13}, we describe the action of symmetry on $e$-particle excitations.  For internal symmetry as we consider here, Ref.~\onlinecite{essin13} considered arbitrary Abelian topological order in $d=2$; the results hold without modification for point-like gauge charge excitations in $d > 2$ topological orders, including the $e$-particles of $\z_n$ topological order.  We assume that symmetry does not permute the different types of anyons, so that symmetry operations take $e$-particles to $e$-particles.  To extract the characteristic symmetry fractionalization of $e$-particles as discussed below, it is sufficient to consider states with $e$-particle excitations only.\cite{essin13}  Therefore, suppose $| \psi \rangle$ is a state with $n$ localized and well-separated $e$-particles.  We expect and assume the property of symmetry localization to hold (see Ref.~\onlinecite{essin13} for further discussion).  That is, 
\begin{equation}
U_g | \psi \rangle = U^e_g(1) \cdots U^e_g(n) | \psi \rangle \text{,}
\end{equation}
where $U^e_g(i)$ is an operator supported in a region localized around the $i$th $e$-particle.  The operators $U^e_g$ (suppressing the $e$-particle label) can be thought of as ``one-particle symmetry operators,'' giving the action of $G$ on a single $e$-particle.  These operators play an important role in the string flux models that are the focus of this paper, not least because $U^e_g$ has a very simple explicit form in these models.

The $U^e_g$ operators form a projective representation,\cite{essin13}
\begin{equation}
U^e_{g_1} U^e_{g_2} = \omega^e (g_1, g_2) U^e_{g_1 g_2} \text{.}
\end{equation}
Here, $\omega^e(g_1, g_2) \in \z_n$ is the $\z_n$ \emph{factor set} characterizing the projective representation. Associativity of $U^e_g$ gives the condition
\begin{equation}
\omega^e (g_1, g_2) \omega^e (g_1 g_2, g_3) = \omega^e(g_2, g_3) \omega^e(g_1, g_2 g_3) \text{.}
\label{eqn:associativity}
\end{equation}
Any function $\omega : G \times G \to \z_n$ satisfying this associativity condition is a $\z_n$ factor set.

We are free to redefine the $U^e_g$ operators by the \emph{projective transformation}
\begin{equation}
U^e_g \to \lambda(g) U^e_g \text{,} \quad \lambda(g) \in \z_n \text{,}
\end{equation}
without affecting the action of symmetry on physical states.  This induces a transformation
on the factor set
\begin{equation}
\omega^e (g_1, g_2) \to \lambda(g_1) \lambda(g_2) \lambda(g_1 g_2)^{-1} \omega^e(g_1, g_2) \text{.}
\label{eqn:projective-transformation}
\end{equation}
Projective transformations reflect an arbitrariness in determining $U^e_g$, and thus bear a family resemblance to gauge transformations in a gauge theory.  Factor sets can be grouped into equivalence classes $[\omega]$ under projective transformations, and the set of these classes is denoted $H^2(G, \z_n)$, which also happens to be the second $\z_n$ cohomology of $G$.  We thus refer to the class $[ \omega ]$ of a factor set $\omega$ as the cohomology class of $\omega$.  $H^2(G, \z_n)$ is an Abelian group, where the group structure comes from the fact that factor sets themselves form an Abelian group, where the product is just multiplication of functions, \emph{e.g.} $\omega_{a b}(g_1, g_2) = \omega_a(g_1, g_2) \omega_b (g_1, g_2)$.  Cohomology classes give a coarser classification of projective representations than that by unitary equivalence; in general, for a given class $[\omega]$, there will be multiple (unitarily inequivalent) projective irreducible representations of $G$.

The $H^2(G, \z_n)$ cohomology classes are physically important, because they uniquely label possible \emph{fractionalization classes} of $e$-particles.\cite{essin13}  In the string flux models, the $e$-particles have fractionalization class $[\omega^e]$, while $m$-particles have trivial fractionalization class $1 \in H^2(G, \z_n)$.  From this and the fusion rules, it follows that the fractionalization class of a general anyon $e^k m^\ell$ is $[\omega^e]^k$. This specifies a \emph{symmetry class}, which is a robust property of a quantum phase of matter, and cannot be changed unless the system undergoes a phase transition, or the $G$ symmetry is broken.  A given symmetry class can comprise more than one distinct phase of matter, but two states in different symmetry classes belong to distinct phases.

It should be noted that two symmetry classes related by a relabeling of anyons are considered equivalent.  The relabeling Eqs.~(\ref{eqn:e-relabel}, \ref{eqn:m-relabel}) has the effect of sending $[\omega^e] \to [\omega^e]^{-1}$, so that two fractionalization classes related in this way give rise to the same symmetry class in the string flux models.  In fact,  $[\omega^e]$ and $[\omega^e]^{-1}$ give rise to the same SET phase in the string flux models, which is established by gauging the $G$ symmetry in Sec.~\ref{sec:gauging}.

\section{Preliminaries}
\label{sec:preliminaries}

\subsection{$\z_n$ quantum double model on a general graph}
\label{sec:anygraph}

The string flux models are $\z_n$ quantum double models ($\z_n$ generalizations of the $\zz$ toric code), defined on special lattices where the group $G$ acts as an internal symmetry.  To prepare for the construction, it will be useful to first define $\z_n$ quantum double models\cite{kitaev03} on a general graph with vertices $v \in V$ and edges (or links) $\ell \in E$.  When we refer to the edge $\ell$, we do so with an orientation.  We denote by $\bar{\ell}$ the same edge $\ell$, but with reversed orientation.  We allow multiple edges to join the same two vertices.  A cycle $c \in C$ is a subset of edges forming a closed loop, and whenever we refer to a cycle we do so with fixed orientation.  We will select a subset of cycles $P \subset C$, so that $P$ forms an elementary set of cycles in a sense described below.  Elements $p \in P$ are referred to as \emph{plaquettes}.

We place a $n$-dimensional Hilbert space on each \emph{edge} of the graph.  Considering first a single fixed edge $\ell$ (with fixed orientation), the Hilbert space of $\ell$ has basis $ \{ \ket{0}, \ket{1} \dots, \ket{n-1} \}$, 
and we introduce (unitary) operators $a_\ell$ and $e_\ell$ defined by
\begin{eqnarray}
a_\ell \ket{k} &=& \exp\Big( \frac{2\pi i k}{n} \Big) \ket{k} \\
e_\ell \ket{k} &=& \ket{k+1} \text{,}
\end{eqnarray}
where $k = 0,\dots, n-1$ and $\ket{n} \equiv \ket{0}$.
We define
\begin{eqnarray}
a_{\bar{\ell}} &\equiv& a^\dagger_\ell \\
e_{\bar{\ell}} &\equiv& e^\dagger_\ell  \text{,}
\end{eqnarray}
and we have the commutation relations
\begin{equation}
a_\ell e_{\ell'} = \left\{ \begin{array}{ll}
e^{2 \pi i / n} e_{\ell'} a_{\ell} & \ell' = \ell \\
e^{-2 \pi i / n } e_{\ell'} a_{\ell} & \ell' = \bar{\ell} \\
e_{\ell'} a_{\ell} & \ell' \neq \ell , \bar{\ell} \text{.}
\end{array} \right.
\end{equation}
Note that for $n=2$ (the $\zz$ toric code), $e_{\bar{\ell}} = e_\ell$, $a_{\bar{\ell}} = a_\ell$, and orientation of edges plays no role.

\begin{figure}[t]
\includegraphics[width=3in]{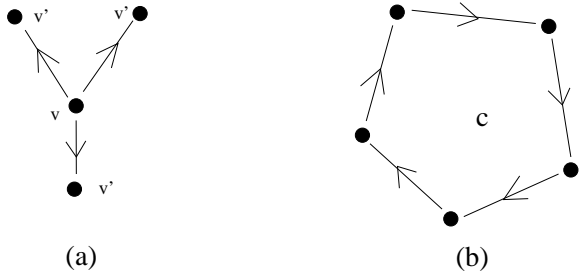}
\caption{(a) Vertices $v'$ adjacent to $v$, and oriented edges used to form $A_v$.  Here and elsewhere, orientation is  denoted by arrows in the middle of an edge.  (b) Edges of an oriented cycle $c$.}
\label{fig:qsbp}
\end{figure}

To define the Hamiltonian, first we need to define operators $A_v$ and $B_p$.  For every vertex $v$ we define
\begin{equation}
A_v \equiv \prod_{\ell \sim v} e_{\ell} \text{,}
\end{equation}
where the product is over all edges $\ell$ joining $v$ to other vertices, with orientation taken pointing away from $v$ (Fig.~\ref{fig:qsbp}a).  For every cycle $c \in C$ we define
\begin{equation}
B_c \equiv \prod_{\ell \in c} a_{\ell} \text{,}
\end{equation}
where the product is taken in an oriented fashion.  It follows from the commutation relations that
\begin{equation}
[A_v, A_{v'}] = [B_c, B_{c'}] = [A_v, B_c] = 0 \text{,}
\end{equation}
for any $v,v' \in V$, $c, c' \in C$.  We will mostly be interested in $B_p$, for plaquettes $p \in P$.

The plaquettes $P$ form an elementary set of cycles in the following sense.  We are interested in graphs that represent $d$-dimensional lattices with periodic boundary conditions, so the $d$-dimensional space in which the system resides is a periodic $d$-dimensional torus $T^d$.  Each cycle $c \in C$ is associated with $d$ integers that characterize its winding around the torus. We say a cycle is contractible if and only if all its winding numbers vanish modulo $n$.  $P$ is chosen so that any contractible cycle can be decomposed into plaquettes.  More precisely, if $c$ is a contractible cycle, we assume that there exist plaquettes $p_1, \dots, p_N \in P$, each taken with some fixed orientation, so that
\begin{equation}
B_c = B_{p_1} B_{p_2} \cdots B_{p_N} \text{.} \label{eqn:cycle-decomposition}
\end{equation}

This definition clarifies why we consider winding numbers modulo $n$ in the definition of contractible cycles.  For example, if the graph is the $d=2$  square lattice, and $P$ consists of the usual square plaquettes, then a cycle winding around the torus $n$ times in one direction is readily decomposed into plaquettes.  Note that if $d=0$, then by definition all cycles are contractible -- we will actually consider a $d=0$ system below in Sec.~\ref{sec:0d} as a building block for the models of interest.

It should be noted that, in general, the set $P$ will be overcomplete.  In particular this means there may be subsets $\{p_1, \dots, p_k \} \in P$ for which
\begin{equation}
B_{p_1} \cdots B_{p_k}  = 1 \text{,}
\end{equation}
so the eigenvalues of $B_p$ cannot all be specified independently.

We consider Hamiltonians of the following form,
\begin{equation}
H = - \sum_{v \in V} (A^{\vphantom\dagger}_v + A^\dagger_v ) - \sum_{p \in P} ( \omega^*_p B_p  + \text{H.c.} ) \text{,}
\label{eqn:generalH}
\end{equation}
where $\omega_p \in \z_n$ has the physical interpretation of a ground-state $\z_n$ flux through plaquette $p$, as we see below.

Since the $A_v$ and $B_p$ operators commute, the model is exactly solvable, and states can be labeled by the eigenvalues of $A_v$ and $B_p$.  In particular, if $| \psi \rangle$ is a ground state, then
\begin{eqnarray}
A_v | \psi \rangle &=&  | \psi \rangle \\
B_p | \psi \rangle &=& \omega_p | \psi \rangle \text{,}
\end{eqnarray}
provided it is actually possible to find a state with these eigenvalues.  This is guaranteed by assuming it is possible to find $| \psi_0 \rangle$, an eigenstate of $a_\ell$ for all $\ell \in E$, satisfying $B_p | \psi_0 \rangle = \omega_p | \psi_0 \rangle$.  Essentially, we are assuming the $B_p$-term of $H$ is un-frustrated.  We can then construct a ground state from $|\psi_0 \rangle$ by writing
\begin{equation}
\ket{\psi_{gs}} = \frac{1}{\sqrt{n}} \prod_{v \in V} \Big[ \frac{1}{\sqrt{n}} \sum_{a = 0}^{n-1} (A_v)^a \Big] \ket{\psi_0} \text{.} \label{eqn:onegs}
\end{equation}
Because $e^n_\ell = 1$, $A^n_v = 1$, so it is easy to see that $A_v \ket{\psi_{gs}} = \ket{\psi_{gs}}$.  Moreover, because $A_v$ and $B_p$ commute, $B_p \ket{\psi_{gs}} = \omega_p \ket{\psi_{gs}}$.  
Therefore, assuming the $B_p$-term is un-frustrated implies that all ground states separately minimize every term in $H$.

When we specialize to $d \geq 2$, an excited state with $k$ $e$-particles at vertex $v$ will be identified by $A_v | \psi \rangle = e^{2 \pi i k / n} | \psi \rangle$.  Equivalently, we say the $\z_n$ charge at $v$ is $k$.  It is useful to introduce $e$-string operators that move such excitations around.  We let $W$ be the set of connected paths in the graph, and $W_o \subset W$ the subset of open paths.  (Cycles are closed paths, so $C \subset W$.)
If $w \in W_o$ is taken with orientation running from the initial endpoint $v_I(w)$ to the final endpoint $v_F(w)$, the string operator
\begin{equation}
S^e(w) = \prod_{\ell \in w} a_{\ell}  \label{eqn:estring}
\end{equation}
decreases the $\z_n$ charge at $v_I$ by one unit, while increasing that at $v_F$ by one unit, thus moving an $e$-particle from $v_I$ to $v_F$.

\subsection{Zero dimensions: quantum double model on the Cayley graph of $G$}
\label{sec:0d}

The basic idea behind the construction of string flux models in $d \geq 1$ is to define the $\z_n$ quantum double model on a graph that is a $d$-dimensional lattice, on which $G$ acts as an internal symmetry.  This is achieved by building a $d$-dimensional lattice out of \emph{Cayley graphs} of the group $G$.  A Cayley graph, described in detail below, gives a representation of the multiplication table of $G$.\cite{lyndon77}  In this section, we describe the quantum double model on a single Cayley graph.  This is a zero-dimensional quantum mechanics problem and not very interesting in its own right, but is the crucial building block for the $d$-dimensional models described in Sec.~\ref{sec:gend}.

\begin{figure}
\includegraphics[width=3.0in]{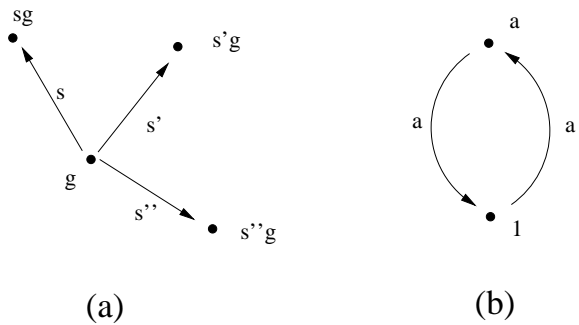}
\caption{(a) Outgoing edges from a vertex $g$ in the Cayley graph of $G$, associated with $s, s', s'' \in S$.  There are also incoming edges (not shown) joining each of $s g, s' g, s'' g$ back to $g$.  The arrows at the end of each edge indicate Cayley graph direction.  (b) Cayley graph of $G = \zz = \{ 1, a \}$.}
\label{fig:cayley}
\end{figure}

A Cayley graph is a directed graph defined given $G$ and a generating set $S \subset G$.  This means that any element of $G$ can be written as a product of elements of $S$.  While other choices are possible  for our purposes, we will always take $S = G \setminus \{1 \}$, and refer to the resulting graph as \emph{the} Cayley graph of $G$. The vertices $v \in V$ are in one-to-one correspondence with group elements $g \in G$, and we can refer to vertices and group elements interchangeably.  Given a vertex $g \in G$, for each $s \in S$ we draw a directed edge joining $g$ to $s g$ (Fig.~\ref{fig:cayley}a).  Every edge is thus associated with left-multiplication by a unique element of $S$.  Given an edge $\ell$, we denote the corresponding group element by $s_\ell \in S$, and we refer to such an edge as a $s_\ell$-edge.  Since $S = G \setminus \{ 1 \}$, every vertex is connected by exactly one outgoing and one incoming edge to every other vertex.  The unique edge joining $g$ to $s_\ell g$, directed away from $g$, can be denoted $\ell = (g, s_\ell g)$.  The Cayley graph of $G = \zz$ is shown in Fig.~\ref{fig:cayley}b.

\begin{figure}
\includegraphics[width=3.0in]{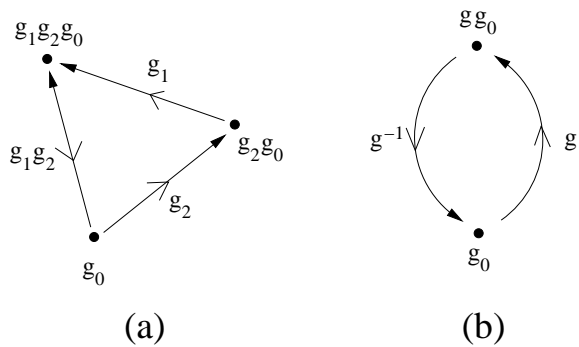}
\caption{Depiction of plaquettes $P$ used to define the quantum double model on a Cayley graph.  (a) General plaquette $p \in P$. We assume $g_1, g_2 \neq 1$, otherwise $g_0, g_1, g_2 \in G$ are arbitrary.   The orientation is taken counterclockwise, as indicated by the arrows in the middle of each edge.  (Recall that the arrows at the end of each edge indicate Cayley graph direction, and not orientation.)  The plaquette in (a) reduces to that shown in (b) if $g_1 g_2 = 1$ and one takes $g_2 = g$ and $g_1 = g^{-1}$.  In this case, the orientation and Cayley graph direction of the edges agree.}
\label{fig:cgcycles}
\end{figure}

In discussing the quantum double model, it is important to distinguish between the \emph{orientation} of an edge and its \emph{Cayley graph direction}.  The direction of each edge is part of the definition of the graph, but we can traverse an edge with arbitrary orientation, either along or against its direction.  We will often need to consider cycles where for some edges the orientation and direction agree, and for other edges they are opposite.  We always indicate orientation by an arrow in the middle of an edge, and direction by an arrow at the end of the edge (see \emph{e.g.} Fig.~\ref{fig:cgcycles}a).

The quantum double model is constructed as for any graph in Sec.~\ref{sec:anygraph}, with the general form of the Hamiltonian given by Eq.~(\ref{eqn:generalH}).  To proceed, we need to specify the plaquettes $P$ and the $\z_n$ fluxes $\omega_p$.  We choose $P$ to consist of cycles of the form shown in Fig.~\ref{fig:cgcycles}a, with $g_1, g_2 \neq 1$.  Note that if $g_2 = g$ and $g_1 = g^{-1}$, the cycle in Fig.~\ref{fig:cgcycles}a reduces to that in Fig.~\ref{fig:cgcycles}b.  Plaquettes can be labeled by the ordered triple  $p = (g_0, g_1, g_2)$, where $g_0$ is arbitrary, and $g_1, g_2 \neq 1$.  In the case $g_1 g_2 \neq 1$, this labeling is unique.  If $g_1 g_2 = 1$ the labeling is not quite unique, as the triples $(g_0, g^{-1}, g)$ and $(g g_0, g, g^{-1})$ both correspond to the cycle shown in Fig.~\ref{fig:cgcycles}b.

\begin{figure}
\includegraphics[width=3in]{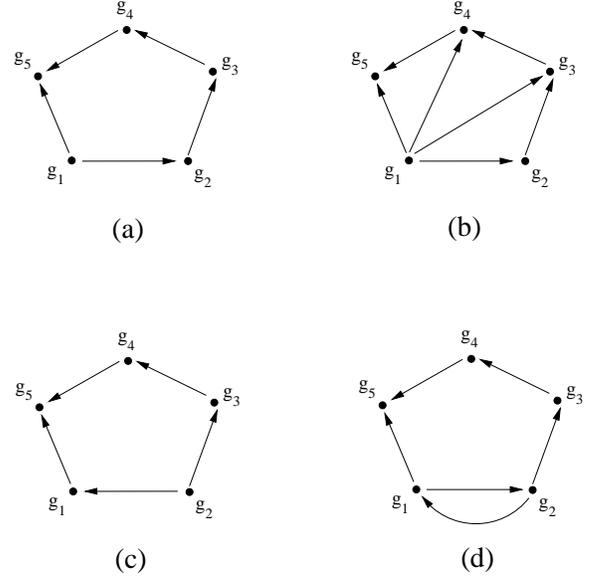}
\caption{Here, we illustrate how to decompose an arbitrary cycle in a Cayley graph into the elementary cycles.  \emph{All} cycles shown are to be traversed with counterclockwise orientation.  In (a), we show a cycle containing five vertices and five edges.  We assume the vertices are all distinct.  The edges are chosen to have Cayley graph direction as shown, pointing from $g_i$ to $g_j$ if $i < j$.  In (b), the cycle of (a) is decomposed into elementary cycles (each triangle).  This procedure generalizes obviously to cycles of arbitrary length, but with the special choice of Cayley graph direction shown.  In (c), we show a cycle with one edge of the ``wrong'' direction.  This is ``repaired'' in (d) by gluing a two-edge elementary cycle as shown.  We can then proceed to apply the procedure depicted in (b), looking only at the ``inner'' edges in (d).}
\label{fig:elementary}
\end{figure}

As required,  any cycle can be decomposed into plaquettes as in Eq.~(\ref{eqn:cycle-decomposition}); a procedure to do this for an arbitrary cycle with non-repeating vertices is given in Fig.~\ref{fig:elementary}.  It is enough to consider cycles with non-repeating vertices, since a cycle with repeating vertices can trivially be decomposed into cycles with non-repeating vertices.

For $p = (g_0, g_1, g_2)$, we choose the flux $\omega_p$ to be
\begin{equation}
\omega_p = \omega(g_1, g_2) \text{,} \label{eqn:flux-choice}
\end{equation}
where $\omega(g_1, g_2)$ is a $\z_n$ factor set of $G$.  The intuition behind this choice is that the plaquette Fig.~\ref{fig:cgcycles}a represents the two ways of multiplying by $g_1 g_2$, either one element at a time, or both together.  The flux through such plaquettes is then naturally associated with a factor set.  Without loss of generality, we require $\omega(g,1) = \omega(1,g) = 1$ for all $g \in G$; this can always be achieved via a suitable projective transformation.  The choice Eq.~(\ref{eqn:flux-choice}) is consistent with the non-uniqueness in labeling because, if $p_1 = (g_0, g^{-1}, g)$ and $p_2 = (g g_0, g, g^{-1})$, then
\begin{equation}
\omega_{p_1} = \omega(g^{-1}, g) = \omega(g, g^{-1} ) = \omega_{p_2} \text{.}
\end{equation}
The middle equality follows from the associativity condition on the factor set [Eq.~(\ref{eqn:associativity})], putting $g_1 \to g$, $g_2 \to g^{-1}$, $g_3 \to g$.

It remains to be checked that this choice of $\omega_p$ leads to an un-frustrated $B_p$-term of the Hamiltonian.  We do this following the discussion of Sec.~\ref{sec:anygraph}, by explicitly constructing $| \psi_0 \rangle$ satisfying $a_\ell | \psi_0 \rangle = a^0_\ell | \psi \rangle$ and $B_p | \psi_0 \rangle = \omega_p | \psi_0 \rangle$.  We choose
\begin{equation}
a^0_{(g_1, g_2 g_1)} = \omega(g_2, g_1) \text{,} \quad g_1, g_2 \in G \text{,} \quad g_2 \neq 1 \text{.}
\end{equation}
Denoting by $B^0_p$ the $B_p$-eigenvalue of $| \psi_0 \rangle$,
for $p = (g_0, g_1, g_2)$ it follows that
\begin{eqnarray}
B^0_p &=& a^0_{(g_0, g_2 g_0)} a^0_{( g_2 g_0, g_1 g_2 g_0) } [ a^0_{(g_0, g_1 g_2 g_0 ) } ]^{-1} \\
&=& \omega(g_2, g_0) \omega(g_1, g_2 g_0) \omega^{-1}(g_1 g_2, g_0) \\
&=& \omega(g_1, g_2) = \omega_p \text{,}
\end{eqnarray}
as desired.  The intermediate steps above do not make sense if $g_1 g_2 = 1$, but the necessary modifications are trivial, and the result $B^0_p = \omega_p$ continues to hold.

We now show that the Hamiltonian has a unitary $G$ symmetry.  We choose $G$ to act on the graph by left-multiplication on vertices, so for the symmetry operation $g \in G$ and the vertex $g_0 \in G$, $g_0 \mapsto g g_0$.  This induces the following action on a $s_\ell$-edge $\ell = (g_0, s_\ell g_0)$:
\begin{equation}
(g_0, s_\ell g_0) \mapsto (g g_0, g s_\ell g_0) = (g g_0, ( g s_\ell g^{-1} ) g g_0 ) \text{.}
\end{equation}
The transformed edge is a $g s_\ell g^{-1}$ edge, so that $G$ acts on $s_\ell$ by conjugation.  Formally, we write the action of $G$ on vertices and edges by $v \mapsto g v$, $\ell \mapsto g \ell$.

Letting $U_g$ be the unitary operator representing $g$, we choose
\begin{equation}
U_g a_{\ell} U^{-1}_g = \Lambda_g(s_\ell) a_{g \ell} , \quad U_g e_{\ell} U^{-1}_g = e_{g \ell} \text{,}
\label{eqn:al-transf}
\end{equation}
where $\Lambda_g(s_\ell) \in \z_n$ is a phase factor depending only on $g$ and on $s_\ell$ and chosen (below) to make $U_g$ a symmetry of $H$, subject to the requirement that $G$ acts linearly on operators in Hilbert space [Eq.~(\ref{eqn:weak-linear-action})].  This requirement gives a condition on $\Lambda_g(s_\ell)$ that can be found by putting ${\cal O} \to a_\ell$ in Eq.~(\ref{eqn:weak-linear-action}), to obtain
\begin{equation}
\Lambda_{g_1} (g_2 s_\ell g_2^{-1} ) \Lambda_{g_2}(s_\ell) = \Lambda_{g_1 g_2} (s_\ell) \text{.} \label{eqn:lambda-condition}
\end{equation}
It turns out that we will also be able to choose $U_g$ to satisfy the stronger condition $U_{g_1} U_{g_2} = U_{g_1 g_2}$.

It is clear that the $A_v$-term of $H$ is invariant under $U_g$.  To understand what happens to the $B_p$-term, and determine $\Lambda_g(s_\ell)$, we consider $p = (g_0, g_1, g_2)$, so that $gp = (g g_0, g g_1 g^{-1}, g g_2 g^{-1} )$.  For $U_g$ to be a symmetry, we need
\begin{equation}
U_g B_p \omega^*_p U^{-1}_g = B_{gp} \omega^*_{gp} \text{.}
\end{equation}
This implies $\Lambda_g(s_\ell)$ must be chosen so that
\begin{equation}
\omega(g_1, g_2) = \Lambda_g(g_1) \Lambda_g(g_2) \Lambda^{-1}_g (g_1 g_2) \omega(g g_1 g^{-1} , g g_2 g^{-1} ) \text{.}
\label{eqn:lambda-symmetry-requirement}
\end{equation}
Interestingly, this is precisely the condition that the two factor sets $\omega(g_1, g_2)$ and $\omega'(g_1, g_2) = \omega(g g_1 g^{-1}, g g_2 g^{-1})$ are equivalent.  This is the case, which we show in Appendix~\ref{app:lambda} by finding $\Lambda_g(s_\ell)$ to be 
\begin{equation}
\Lambda_g(s_\ell) = \omega^{-1}(g^{-1}, g s_\ell ) \omega(g s_\ell, g^{-1} ) \text{.} \label{eqn:lambda}
\end{equation}
It is also shown in Appendix~\ref{app:lambda} that $\Lambda_g(s_\ell)$ indeed satisfies Eq.~(\ref{eqn:lambda-condition}), and that $U_g$ can be chosen to satisfy $U_{g_1} U_{g_2} = U_{g_1 g_2}$.

\section{String flux models with $G$ symmetry and $\z_n$ topological order}
\label{sec:gend}

\begin{figure}
\includegraphics[width=0.8\columnwidth]{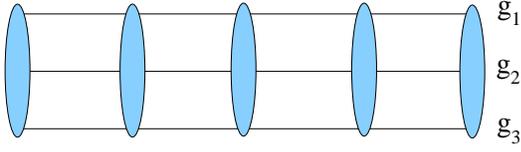}
\caption{Lattice for the $d=1$ string flux model, visualized as a stack of $|G|$ one-dimensional chains, each labeled by a group element $g \in G$.  The shaded ovals represent the Cayley subgraphs that connect the chains together.}
\label{fig:1d}
\end{figure}

We are now prepared to describe the models of interest.  We consider $|G|$ copies of a $d$-dimensional hypercubic lattice, with each copy labeled by a group element $g \in G$ and referred to as the $g$-hypercubic lattice.\footnote{Changing the lattice requires only trivial modifications. We choose hypercubic for ease of visualization and ease of discussing all dimensions $d$ simultaneously.}   In $d=1$ this is conveniently visualized as a stack of $|G|$ chains (Fig.~\ref{fig:1d}), and in $d=2$ as a multilayer square lattice with $|G|$ layers (Fig.~\ref{fig:mparticle}a).  Hypercubic lattice primitive cells are labeled by $\br = (n_1, n_2, \dots, n_d) \in \z^d$.  The vertex with spatial position $\br$ in the $g$-hypercubic lattice is denoted $v(\br, g)$.  The set of edges of the $g$-hypercubic lattice is denoted $E_S(g)$, where the subscript stands for ``spatial.''  The set of all spatial edges is $E_S$.  $\ell \in E_S(g)$ is referred to as a $g$-edge, or more specifically as a spatial $g$-edge.

So far, we have not described how to connect the $|G|$ copies of the lattice together.  This is done by connecting the vertices in each primitive cell as a Cayley graph.  We denote the set of Cayley graph edges  (Cayley edges for short) in primitive cell $\br$ by $E_C(\br)$, and the set of all Cayley edges by $E_C$.  An edge $\ell \in E_C$ can be denoted uniquely by $\ell = (\br ; g_0, s_\ell g_0)$, where $g_0 \in G$, $s_\ell \in S = G \setminus \{1 \}$.  As in Sec.~\ref{sec:0d}, such an edge is referred to as a $s_\ell$-edge, or more specifically as a Cayley $s_\ell$-edge.  We see that every edge of the lattice is associated with a group element, but it should be kept in mind that the meaning of this association is different for Cayley and spatial edges.

\begin{figure}
\includegraphics[width=\columnwidth]{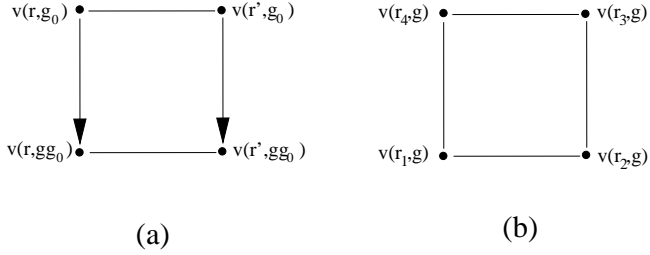}
\caption{(a) General type II plaquette $p \in P_2$. $\br, \br'$ are neighboring hypercubic lattice sites, and $g_0, g \in G$, with $g \neq 1$.  Cayley graph direction is indicated on the two Cayley edges, while spatial edges are undirected.  (b) General type III plaquette $p \in P_3$.  $\br_1, \dots, \br_4$ are hypercubic lattice sites forming a square face, and $g \in G$.  We will not need to specify a conventional orientation for type II and type III plaquettes, so the plaquettes here are drawn without orientation.}
\label{fig:type2-type3}
\end{figure}

Having defined the lattice, we place spins on the edges and construct the quantum double model as for a general graph (Sec.~\ref{sec:anygraph}).  The Hamiltonian again takes the form given in Eq.~(\ref{eqn:generalH}), and to specify it we need to define plaquettes $P$ and $\z_n$ fluxes $\omega_p$.

We divide the plaquettes into three types, writing $P =  P_1 \cup P_2 \cup P_3$.  Type I plaquettes $P_1$ comprise the plaquettes introduced in Sec.~\ref{sec:0d}, for each Cayley subgraph.  Type II plaquettes $P_2$ connect two nearest-neighbor Cayley subgraphs, whose primitive cells are nearest-neighbor hypercubic lattice sites $\br, \br'$.  Elements $p \in P_2$ are uniquely specified by  $\br, \br'$ and group elements $g_0, g \in G$ (with $g \neq 1$); the corresponding plaquette is shown in Fig.~\ref{fig:type2-type3}a.  Type III plaquettes $P_3$, which are only present for $d \geq 2$, are simply the square faces of each $g$-hypercubic lattice (see Fig.~\ref{fig:type2-type3}b). $p \in P_3$ is uniquely specified by a face of the hypercubic lattice (containing sites $\br_1, \br_2, \br_3, \br_4$) and a group element $g$.  $p$ lies in the $g$-hypercubic lattice, joining the four vertices $v(\br_1, g), v(\br_2, g), v(\br_3, g), v(\br_4, g)$.

\begin{figure}
\includegraphics[width=\columnwidth]{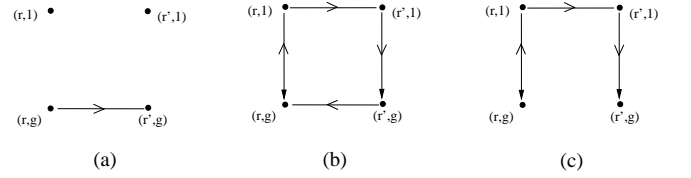}
\caption{(a) Spatial $g$-edge in a cycle $c$, connecting hypercubic sites $\br$ and $\br'$.  (b) A type II plaquette.  (c) Result of gluing together the edge in (a) and the plaquette in (b).  The effect is to ``move over'' the segment of $c$ shown in (a) to become a spatial $1$-edge, at the expense of creating  additional edges within the Cayley subgraphs at $\br$ and $\br'$.}
\label{fig:moving_g-edge}
\end{figure}

Before proceeding, we give a procedure to decompose an arbitrary contractible cycle $c$ into plaquettes.  We can view such a decomposition as a procedure to contract the cycle down to nothing, by successively ``gluing'' plaquettes.  First, suppose $c$ contains a spatial $g$-edge joining $\br$ to $\br'$.  We can glue a type II plaquette, so that this $g$-edge becomes a $1$-edge connecting the same two hypercubic sites, as shown in Fig.~\ref{fig:moving_g-edge}. This process adds new edges to $c$ that lie within the Cayley subgraphs.  We can then proceed this way until all edges in $c$ lie within a Cayley subgraph, or are spatial $1$-edges.  
Next, note that $c$ now both enters and exits a given Cayley subgraph at the $g = 1$ vertex.  This gives a closed loop within the Cayley subgraph, which can be contracted using type I plaquettes.  This allows us to eliminate all edges within Cayley subgraphs.  We are left with a cycle consisting only of spatial $1$-edges, which is just a cycle in the $d$-dimensional hypercubic lattice.  As long as this cycle is contractible, it can be contracted using type III plaquettes.

We can denote $p \in P_1$ by $(\br; g_0, g_1, g_2)$ (see Sec.~\ref{sec:0d}), and we choose choose $\omega_p = \omega(g_1, g_2)$, independent of $\br$.  For $p \in P_2, P_3$, we choose simply $\omega_p = 1$.  The latter choice means we will not need to choose a conventional orientation for type II and III plaquettes, in contrast to type I plaquettes.  To show that the $B_p$-term is unfrustrated for this choice of fluxes, we let $|\psi_0 \rangle$ be an eigenstate of $a_\ell$ for all $\ell \in E$, with $a^0_\ell$ the eigenvalue of $a_\ell$, and
\begin{equation}
a^0_\ell = 
\begin{cases}
1 , &  \ell \in E_S \\
\omega(g_2, g_1) ,  &  \ell = (\br ; g_1, g_2 g_1) \in E_C
\end{cases} \text{.}
\end{equation}
It is easily seen that $B_p | \psi_0 \rangle = \omega_p | \psi_0 \rangle$.

The action of $G$ on the lattice is defined by its action on vertices, which is in turn given by left-multiplication in each Cayley subgraph.  That is, for $g \in G$,
\begin{equation}
v(\br, g_0) \mapsto v(\br, g g_0) \text{.}
\end{equation}
$G$ thus acts on spatial edges by left-multiplication.  That is, $g \in G$ takes the $g_0$-edge joining $\br$ and $\br'$ to the $g g_0$-edge joining the same hypercubic sites.  The action on a Cayley edge $\ell = (\br ; g_1, g_2 g_1)$ is
\begin{equation}
 (\br ; g_1, g_2 g_1) \mapsto (\br ; g g_1, g g_2 g_1) \text{.}
\end{equation}
As before, formally we write $v \mapsto g v$, $\ell \mapsto g \ell$.  The symmetry acts on operators by
\begin{eqnarray}
U_g e_\ell U^{-1}_g &=& e_{g \ell} , \quad \ell \in E \\
U_g a_\ell U^{-1}_g &=& a_{g \ell} , \quad \ell \in E_S \\
U_g a_\ell U^{-1}_g &=& \Lambda_g(s_\ell) a_{g \ell} , \quad \ell \in E_C \text{,}
\end{eqnarray}
with $\Lambda_g(s_\ell)$ given by Eq.~(\ref{eqn:lambda}).  $U_g$ thus defined can be chosen to satisfy $U_{g_1} U_{g_2} = U_{g_1 g_2}$, which follows immediately from the fact that this relation holds for a single Cayley graph (Sec.~\ref{sec:0d} and Appendix~\ref{app:lambda}).

\section{Topological order of the string flux models}
\label{sec:topo}

For $d \geq 2$, the string flux models have $\z_n$ topological order.  That is, they have the topological order of the $\z_n$ quantum double model on the $d$-dimensional hypercubic lattice.  Put another way, going from the simple $d$-dimensional hypercubic lattice, to the rather intricate lattices of Sec.~\ref{sec:gend}, does not alter the topological order.  We shall focus on $d = 2$ and briefly describe the generalization to $d > 2$ at the end of the section.

First, we observe that the ground state degeneracy on a torus is exactly $n^2$.  This can be seen via the construction of Sec.~\ref{sec:gmapping}, where the coefficients of type I and II plaquette terms are taken large, and the resulting low-energy limit is mapped onto a $\z_n$ gauge theory with bosonic matter on the $d = 2$ square lattice.  This theory is well known to have a $n^2$ ground state degeneracy.

In fact, the mapping to $\z_n$ gauge theory not only determines the ground state degeneracy, but completely establishes the presence of $\z_n$ topological order.  However, to study the role of symmetry, it will be useful to proceed without mapping to gauge theory.  We explicitly identify the $\z_n$ charge and flux excitations ($e$ and $m$ particles), and the associated string operators that create and move these excitations.  The topological order is determined by the algebraic properties of these string operators.

$e$-particles reside at vertices.  If  a state satisfies $A_v | \psi \rangle = e^{ 2 \pi i k / n} | \psi \rangle$, we say there are $k$ $e$-particles at $v$ (equivalently, $\z_n$ charge of $k$ at $v$).  The string operator $S^e(w)$ moving an $e$-particle from $v_I$ to $v_F$, where $w \in W_o$ is an open path with endpoints $v_I$ and $v_F$, is $S^e(w) = \prod_{\ell \in w} a_\ell$ [Eq.~(\ref{eqn:estring})].  Acting on a ground state with $S^e(w)$ creates an excitation with an $e$-particle at $v_F$, and $(n-1)$ $e$-particles at $v_I$.  Equivalently, we can say that an anti-$e$ or $\bar{e}$ particle is created at $v_I$.  (Note that $\bar{e} = e^{n-1}$.)  Any state with a multiple of $n$ $e$-particles can be created by acting with an appropriate product of such operators.

\begin{figure}
\includegraphics[width=\columnwidth]{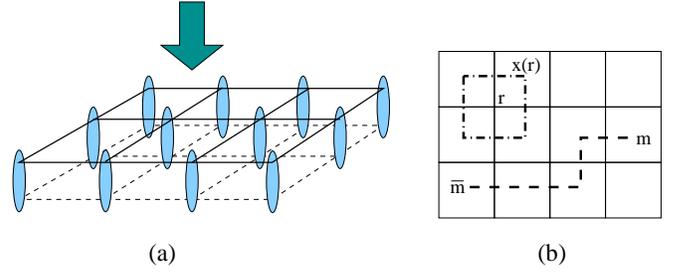}
\caption{(a) $d=2$ model visualized as a vertical stack of square lattices.  The top layer is indicated by solid lines, and the bottom layer by dashed lines (other layers in between not shown).  Shaded ovals represent the Cayley graphs connecting the layers, and the shaded arrow indicated that we view the stack of square lattices from above to obtain the planar projection.  (b) Shows the resulting planar projection, which is simply a square lattice, with a cut (dashed line).  The cut has endpoints in plaquettes of the square lattice, and intersects square edges, but not vertices.  $m$ and $\bar{m}$ particles lie at the endpoints of the cut.  The dash-dot line shows a closed cut $x(\br)$, encircling the single square lattice site $\br$, which is used in Sec.~\ref{sec:gmapping}.}
\label{fig:mparticle}
\end{figure}

To identify $m$-particles, it is helpful to visualize the lattice as a vertical stack of $|G|$ square lattices (Fig.~\ref{fig:mparticle}a).  Viewing this stack from above defines a projection to a single square lattice in the two-dimensional plane.  Each edge of this projected square lattice represents $|G|$ different spatial edges.  All the vertices and edges in the Cayley graph at $\br$ project to a single point in the plane.  In the planar projection, $m$-particles reside at square plaquettes, and are created by threading one unit of $\z_n$ flux vertically through the stack of square lattices.  $m$-string operators are defined on \emph{cuts} $x \in X$ as illustrated in Fig.~\ref{fig:mparticle}b, and
\begin{equation}
S^m(x) = \prod_{\ell \in x} e_{\ell} \text{.}
\end{equation}
The edges $\ell \in x$ are those spatial edges that cross the cut $x$ under planar projection.  This operator increases the $\z_n$ flux by one unit at one endpoint, creating an $m$-particle there.  At the other endpoint it decreases the flux by one unit, creating an anti-$m$ or $\bar{m}$-particle.

Now we turn to the algebraic properties of the string operators.  $e$-strings commute with other $e$-strings, and $m$-strings with other $m$-strings.  From this property, the exchange statistics of $e$ and $m$ particles can be computed, and both particles are bosons.  In addition, $[S^e(w)]^n = [S^m(x)]^n = 1$ for any path $w \in W$ and any cut $x \in X$, which implies the fusion rules $e^n = m^n = 1$.  Now, suppose a path $w$ and a cut $x$ cross each once under planar projection, then
\begin{equation}
S^e(w) S^m(x) = \exp \Big( \frac{\pm  2 \pi i}{n} \Big) S^m(x) S^e(w) \text{,} \label{eqn:string-comm-reln}
\end{equation}
where the sign in the exponential depends on the orientation of the crossing.  This property is responsible for the $\theta = 2\pi / n$ mutual statistics of $e$ and $m$ particles.

For large, closed paths $w$ and cuts $x$ that wind around the torus, the algebraic properties of the string operators  imply that the ground state degeneracy is at least $n^2$.  Since the degeneracy is exactly $n^2$, we have not missed anyons beyond those generated by fusion of $e$ and $m$; if present, such particles would result in a larger ground state degeneracy.

This discussion generalizes in a straightforward fashion to dimensions $d > 2$.  The mapping to $\z_n$ gauge theory (Sec.~\ref{sec:gauging})) holds in any dimension, and the ground state degeneracy is $n^d$.  $e$-particles again reside at vertices, and $e$-string operators are defined just as above.  The $d=2$ planar projection generalizes naturally to a projection to  $d$-dimensional space.  $\z_n$ flux excitations are now $(d-2)$-dimensional objects.  In $d=3$, these are flux lines, where $\z_n$ flux is threaded through the cubic lattice (under projection).  The generalization of the $m$-string operator is a $(d-1)$-dimensional membrane operator, which is a product of $e_\ell$ over edges $\ell$ crossed by the membrane.  Commutation of string and membrane operators results in a phase factor $e^{\pm 2\pi i / n}$ for each time the string pierces the membrane, generalizing Eq.~(\ref{eqn:string-comm-reln}).  These properties establish the presence of $\z_n$ topological order.

\section{Symmetry fractionalization in $d \geq 2$}
\label{sec:sf2d}

We now study symmetry fractionalization of $G$ in the string flux models for dimensions $d \geq 2$.  We first discuss the general properties of the one-particle symmetry operators $U^e_g$.  We show that a choice of $U^e_g$ is unique up to $\z_n$ projective transformations and forms a projective representation with a well-defined $\z_n$ factor set.  We then find the explicit form of $U^e_g$, and show that the $e$-particle fractionalization class $[\omega^e]$ is given by $[\omega^e] = [\omega] \in H^2(G, \z_n)$, where $\omega(g_1, g_2) = \omega_p$ is the factor set directly encoded into the Hamiltonian as a pattern of ground-state $\z_n$ fluxes through type I plaquettes.  Some technical details are given in Appendix~\ref{app:eloc}.

To construct $U^e_g$ operators  and extract the corresponding $H^2(G, \z_n)$ fractionalization class,  it will be enough to consider states with $n$ $e$-particles and no other excitations (see Sec.~\ref{sec:background}).  We denote such a state by $|\psi_e \rangle$, with $e$-particles at vertices $v_1, \dots, v_n$.  We would then like to find operators satisfying
\begin{equation}
U_g | \psi_e \rangle = U^e_g(v_1) \cdots U^e_g(v_n) | \psi_e \rangle \text{.}
\label{eqn:eloc-general}
\end{equation}
As suggested by the notation, the operators $U^e_g(v)$ depend only on $g \in G$ and on the vertex at which an $e$-particle resides.  Because the state of a single $e$-particle is completely specified by its vertex, this is the only information the $U^e_g$'s may depend on if they are to give a localization of the symmetry.  For the same reason, given a choice of $U^e_g(v)$ for all $g \in G$ and $v \in V$, we demand that this choice satisfy Eq.~(\ref{eqn:eloc-general}) for any state $|\psi_e\rangle$ with $n$ $e$-particles.  This also ensures that the obvious generalization of Eq.~(\ref{eqn:eloc-general}) holds for states with larger numbers of $e$-particles.

It is clear that $U^e_g(v)$ must move an $e$-particle from $v$ to $g v$. Therefore, $U^e_g(v)$ must be proportional to an $e$-string operator on a path joining $v$ to $g v$.  Since we restrict to states where the only excitations are $e$-particles, deforming the path while keeping the endpoints fixed is equivalent to multiplying $U^e_g(v)$ by a $\z_n$ phase factor.  This means that the choice of $U^e_g(v)$ is at most arbitrary under transformations $U^e_g(v) \to \lambda(g, v) U^e_g(v)$, where $\lambda(g,v) \in {\rm U}(1)$.  However, the requirement that Eq.~(\ref{eqn:eloc-general}) holds for any choice of $n$ $e$-particles implies that $\lambda(g,v) = \lambda(g)$, independent of $v$, and moreover $\lambda(g) \in \z_n$.\footnote{This can be seen by considering the $n+1$ states formed by placing $e$-particles at $n$ of the vertices $v_1, \dots, v_{n+1}$.}  Therefore, the choice of $U^e_g(v)$ is unique up to $\z_n$ projective transformations  $U^e_g(v) \to \lambda(g) U^e_g(v)$ ($\lambda(g) \in \z_n$).

We expect that $U^e_g(v)$ gives a projective representation with a well-defined $\z_n$ factor set.  This can be shown by considering the product of two successive $U^e_g(v)$'s,
\begin{equation}
U^e_{g_1}(g_2 v) U^e_{g_2}(v) | \psi_e \rangle = \omega^e_v(g_1, g_2) U^e_{g_1 g_2} (v)  | \psi_e \rangle \text{,}
\end{equation}
where $| \psi_e \rangle$ has an $e$-particle at $v$, and $\omega^e_v(g_1, g_2) \in {\rm U}(1)$. The equation follows from the fact that the operators on the left- and right-hand sides both move an $e$-particle from $v$ to $g_1 g_2 v$, and thus must be proportional.  In fact, Eq.~(\ref{eqn:eloc-general}) together with $U_{g_1} U_{g_2} = U_{g_1 g_2}$ implies $\omega^e_v(g_1, g_2) = \omega^e(g_1, g_2) \in \z_n$.  Moreover, $\omega^e(g_1, g_2)$ is a $\z_n$ factor set, because multiplication of $U^e_g$'s is associative.  This factor set is well-defined up to projective transformations, so $U^e_g(v)$ uniquely determines the $e$-particle fractionalization class $[ \omega^e ] \in H^2(G, \z_n)$.

Now we proceed by writing down a choice for $U^e_g(v)$.  We claim that
\begin{equation}
U^e_g[ v(\br, g_0) ] =  \left\{  \begin{array}{ll} 
a_{(\br ; g_0, g g_0)}   & , \quad g \neq 1 \\
1  & , \quad g = 1 \end{array}\right. \text{.} \label{eqn:ueg}
\end{equation}
It should be recalled that $\ell = (\br; g_0, g g_0)$ is a Cayley $g$-edge in the Cayley subgraph at $\br$.  It is clear that this choice moves an $e$-particle from $v(\br, g_0)$ to $g v = v(\br, g g_0)$.  A direct calculation in Appendix~\ref{app:eloc} shows that this choice of $U^e_g(v)$ satisfies Eq.~(\ref{eqn:eloc-general}) for all states $|\psi_e\rangle$ with $n$ $e$-particles.

To extract the fractionalization class $[\omega^e]$, we focus on an $e$-particle at vertex $v= v(\br, g)$ in $| \psi_e \rangle$.  We consider the type I plaquette $p = (\br; g, g_1, g_2)$, and note that
\begin{equation}
B_p = U^e_{g_1}(g_2 v) U^e_{g_2}(v) [ U^e_{g_1 g_2} (v) ]^{-1} \text{.}
\end{equation}
Since $B_p | \psi_e \rangle = \omega_p | \psi_e \rangle$, where $\omega_p = \omega(g_1, g_2)$, we have
\begin{equation}
U^e_{g_1}(g_2 v) U^e_{g_2}(v) | \psi_e \rangle = \omega(g_1, g_2) U^e_{g_1 g_2} (v)  | \psi_e \rangle \text{.}
\end{equation}
Therefore, we have shown $\omega^e(g_1, g_2) = \omega(g_1, g_2)$, and
\begin{equation}
[\omega^e] = [ \omega ] \text{.}
\end{equation}
We have thus established one of our main results, namely that our models can realize any $e$-particle fractionalization class $[\omega^e]$, since $\omega$ is an arbitrary $\z_n$ factor set.  In addition, $[\omega^e]$ is directly encoded into the Hamiltonian via the ground-state $\z_n$ fluxes $\omega_p$.

\section{Mapping to $\z_n$ gauge theory and magnetic route to low-energy gauge theory}
\label{sec:gmapping}

The string flux models map to $\z_n$ gauge theories with bosonic matter in a suitable low-energy limit.  The mapping to gauge theory is particularly useful as a route to study the result of gauging $G$ symmetry in Sec.~\ref{sec:gauging}.  The mapping is also interesting in its own right, as a novel \emph{magnetic} route to obtain a low-energy effective gauge theory.  There are many simple models in which gauge theory emerges at low energy via an \emph{electric} mechanism, due to the presence of a large term in the Hamiltonian enforcing the Gauss' law constraint.\cite{kitaev03,balents02,senthil02,motrunich02,wen03b,hermele04,levin05}  Here, the emergence of gauge theory has to do with constraints on the magnetic ($m$-particle) excitations.  The electric and magnetic routes to low-energy gauge theory may be dual in some sense.

The result holds in any dimension, but, to simplify the discussion, we focus on $d=2$, where we can visualize the lattice as a multilayer of $|G|$ square lattices, with a planar projection to a single square lattice (see Sec.~\ref{sec:topo} and Fig.~\ref{fig:mparticle}a).  We modify the Hamiltonian as follows:
\begin{equation}
H = - \sum_{v \in V} (A^{\vphantom\dagger}_v + A^\dagger_v ) - \sum_{p \in P} K_p ( \omega^*_p B_p  + \text{H.c.} ) \text{.}
\end{equation}
Here, we have introduced the real parameter $K_p > 0$.  We choose $K_p = 1$ for type III plaquettes, and take the limit $K_p \to + \infty$ for type I and II plaquettes.  Taking this limit imposes the constraint $B_p = \omega_p$ for $p \in P_1, P_2$, defining a low-energy Hilbert space that we now study.  As we see below, this limit retains all the distinct types of anyons present in $\z_n$ topological order.

Excitations in the low-energy Hilbert space are precisely the $e$ and $m$ particles described in Sec.~\ref{sec:topo}.  While arbitrary configurations of $\z_n$ charge at vertices are allowed, the \emph{only} $\z_n$ flux excitations are obtained by threading flux vertically through the stack of square lattices, so that flux only passes through type III plaquettes.  Any other configurations of $\z_n$ flux are outside the low-energy Hilbert space.  This means that, if $p$ and $p'$ are two type III plaquettes identified under planar projection, $B_p = B_{p'}$ in the low-energy Hilbert space.  It is thus convenient to define $P_s \subset P_3$ to be the set of type III plaquettes in the $1$-square lattice (\emph{i.e.}, type III plaquettes in the $g=1$ layer only).  A complete basis of energy eigenstates for the low-energy Hilbert space is provided by the simultaneous eigenstates of the operators $\{ A_v | v \in V \}$, $\{ B_p | p \in P_s \}$, $B_{c_x}$, and $B_{c_y}$.  Here, $c_x$ ($c_y$) is a large cycle winding around the torus once in the $x$ ($y$) direction.

We now construct a $\z_n$ gauge theory on the square lattice, with precisely the same excitations and (gauge-invariant) Hilbert space.  We consider the (single-layer) square lattice, with sites $\br$.  As in the construction of the quantum double model, we place a $n$-dimensional Hilbert space on each nearest-neighbor edge $\br \br'$, and introduce $\z_n$ vector potential ($\cA_{\br \br'}$) and electric field ($\cE_{\br \br'}$) operators acting on the edge degrees of freedom.  $\cA_{\br \br'}$ and $\cE_{\br \br'}$ satisfy the same properties as $a_\ell$ and $e_\ell$ in the quantum double model (see Sec.~\ref{sec:anygraph}).  On each site $\br$, we introduce a Hilbert space with basis $| \{ N_g | g \in G \} \rangle$, where $N_g \in \{ 0, \dots, n-1\}$, suppressing site labels for the moment.  We introduce number operators $\hat{N}_g$ defined by
\begin{equation}
\hat{N}_g | \{ N_{g'} | g' \in G \} \rangle = N_g | \{ N_{g'} | g' \in G \} \rangle \text{,}
\end{equation}
and raising (lowering) operators $b^\dagger_g$ ($b_g$), defined by
\begin{eqnarray}
b^\dagger_g | \dots, N_g , \dots \rangle &=&   |  \dots, N_g + 1, \dots \rangle \\
b_g |  \dots, N_g , \dots \rangle &=&   |  \dots, N_g - 1, \dots \rangle \text{,}
\end{eqnarray}
where $N_g + 1$ and $N_g - 1$ are to be interpreted modulo $n$.
It should be noted that $b_g, b^\dagger_g$ are not canonical boson operators.  Restoring site labels, we denote these operators by $\hat{N}_{\br g}$, $b^\dagger_{\br g}$, and $b_{\br g}$.

We would like $b^\dagger_{\br g}$ to create an $e$-particle excitation with unit $\z_n$ gauge charge at site $\br$, so we impose the gauge constraint
\begin{equation}
\prod_{\br' \sim \br} \cE_{\br \br'} = \exp \Big( \frac{2 \pi i}{n} \sum_{g \in G} \hat{N}_{\br g}  \Big) \text{,}
\end{equation}
where the product is over the four sites $\br'$ adjacent to $\br$.  We consider the Hamiltonian
\begin{equation}
H_{{\rm gauge}} = - 2 \sum_{\br} \sum_{g \in G} \cos \Big( \frac{2\pi}{n} \hat{N}_{\br g} \Big) 
- n \sum_{p \in P_s} \big[ \cB_p + \text{H.c.} \big] \text{,}
\label{eqn:Hgauge}
\end{equation}
where $\cB_p = \prod_{\br \br' \in p} \cA_{\br \br'}$.
This Hamiltonian is a sum of mutually commuting terms, and is exactly solvable.  A more generic $\z_n$ gauge theory would have an electric-field term.  Here, this term has been set to zero, putting the model deep into the deconfined phase.  In a ground state, $\hat{N}_{\br g} = 0$ and $\cB_p = 1$.  More generally, a complete basis of energy eigenstates is provided by simultaneous eigenstates of $\{ \hat{N}_{\br g} \}$, $\{ \cB_p \}$, $\cB_{c_x}$, and $\cB_{c_y}$, where $c_x$ and $c_y$ are large cycles as above.  This is the same Hilbert space as obtained in the low-energy limit of the quantum double model; the correspondence between states is made clear by the discussion below.

To complete the mapping between the gauge theory and the quantum double model, we now establish an operator dictionary relating these two models.  We begin with the operators appearing in the Hamiltonian, for which we have
\begin{eqnarray}
A_{v(\br, g)} &\leftrightarrow& \exp\Big( \frac{2 \pi i}{n} \hat{N}_{\br g} \Big) \label{eqn:av-corr} \\
B_p &\leftrightarrow& \cB_p , \quad ( p \in P_s ) \text{.}
\end{eqnarray}
This correspondence tells us how to relate energy eigenstates in the two models, which (by deliberate construction) have identical energies.

We also need to relate $e$ and $m$ string operators in the two models.  As discussed in Sec.~\ref{sec:topo}, $m$-strings in the quantum double model are defined on cuts $x \in X$, and $S^m(x) = \prod_{\ell \in x} e_{\ell}$.  In the gauge theory, $m$-strings are given very similarly by ${\cal S}^m(x) = \prod_{\br \br' \in x} \cE_{\br \br'}$, and we have
\begin{equation}
S^m(x) \leftrightarrow {\cal S}^m(x) \text{.} \label{eqn:ms-corr}
\end{equation}
If we substitute Eqs.~(\ref{eqn:av-corr}) and~(\ref{eqn:ms-corr}) into the gauge constraint, we obtain a relation for quantum double model operators,
\begin{equation}
\prod_{g \in G} A_{v(\br, g)} = S^m[ x(\br) ] \text{,}
\end{equation}
where $x(\br)$ is the closed cut enclosing only the lattice site $\br$, as shown in Fig.~\ref{fig:mparticle}b.  This relation is easily seen to hold in the quantum double model, so the operator dictionary is compatible with the gauge constraint.

 $e$ string operators in the quantum double model are products of $a_\ell$, so it is enough to add $a_\ell$ to the dictionary.  We do this separately for Cayley and spatial edges.  For a Cayley edge $\ell = (\br; g_0, g g_0)$, $a_\ell$ creates an $e$-particle at $v(\br, g g_0)$, while destroying one at $v(\br, g_0)$.  This motivates the correspondence
 \begin{equation}
a_{(\br; g_0 , g g_0)} \leftrightarrow \omega(g, g_0) b^\dagger_{\br, g g_0} b^{\vphantom\dagger}_{\br g_0} \text{,}  \label{eqn:al-cayley-corr}
\end{equation}
where the phase factor $\omega(g, g_0)$ is needed for consistency with the constraint $B_p = \omega_p$, when $p$ is a type I plaquette.

Now, for the spatial $g$-edge $\ell = (\br, \br' ; g)$, $a_\ell$ creates an $e$-particle at $v(\br', g)$ and destroys one at $v(\br, g)$.
We have
\begin{equation}
a_{(\br, \br'; g)} \leftrightarrow b_{\br g} \cA_{\br \br'} b^\dagger_{\br' g} \text{,} \label{eqn:al-spatial-corr}
\end{equation}
where the gauge field $\cA_{\br \br'}$ appears to make the operator gauge invariant.
From Eqs.~(\ref{eqn:al-cayley-corr}) and~(\ref{eqn:al-spatial-corr}), for any quantum double model $e$-string operator $S^e(w) = \prod_{\ell \in w} a_{\ell}$, we can easily find the corresponding gauge theory operator.

With the operator dictionary complete, we now determine the action of symmetry on the gauge theory degrees of freedom.  The action of symmetry on gauge-invariant operators follows directly from the dictionary.  However, it will be useful to go farther, and extend $U_g$ to act on non-gauge-invariant operators, which is a well-defined problem up to $\z_n$ gauge transformations.  We define
\begin{eqnarray}
U_g b^\dagger_{\br g_0} U^{-1}_g &=& \omega(g, g_0) b^\dagger_{g g_0} \label{eqn:gauge-ug} \\
U_g \hat{N}_{\br g_0} U^{-1}_g &=& \hat{N}_{\br, g g_0} \\
U_g \cA_{\br \br'} U^{-1}_g &=& \cA_{\br \br'} \\
U_g \cE_{\br \br'} U^{-1}_g &=& \cE_{\br \br'} \text{.}
\end{eqnarray}
These definitions reproduce the correct action of symmetry on gauge-invariant operators.  This is verified for $a_\ell$ with $\ell$ a Cayley edge in Appendix~\ref{app:ug-gauge}; verification for other operators is straightforward.

The operator $b^\dagger_{\br g_0}$ creates an $e$-particle ($\z_n$ charge), so we expect it to transform as a projective representation of $G$ in fractionalization class $[\omega^e] = [\omega]$.  This is verified by computing
\begin{equation}
U_{g_1} U_{g_2} b^\dagger_{\br g_0} U^{-1}_{g_2} U^{-1}_{g_1} = \omega(g_1, g_2) U_{g_1 g_2} b^\dagger_{\br g_0} U^{-1}_{g_1 g_2} \text{.}
\end{equation}
Because $U_g$ is only uniquely defined up to $\z_n$ gauge transformations, it is important to show that the fractionalization class is well-defined.  A gauge transformation of $U_g$ is the modification $U_g \to U'_g = G_g U_g$, where
\begin{eqnarray}
G_g b^\dagger_{\br g_0} G^{-1}_{g} &=& \lambda_{\br}(g) b^\dagger_{\br g_0} \\
G_g \cA_{\br \br'} G^{-1}_g &=& \lambda_{\br}(g) \cA_{\br \br'} \lambda^{-1}_{\br'}(g) \text{,}
\end{eqnarray}
for some $\lambda_{\br}(g) \in \z_n$.  This implies
\begin{equation}
U'_g b^\dagger_{\br g_0} (U'_g)^{-1} = \lambda_{\br}(g) \omega(g, g_0) b^\dagger_{\br g_0} \text{,}
\end{equation}
and therefore
\begin{equation}
U'_{g_1} U'_{g_2} b^\dagger_{\br g_0} (U'_{g_2})^{-1} (U'_{g_1})^{-1} =  \omega'_{\br}(g_1, g_2) U'_{g_1 g_2} b^\dagger_{\br g_0} (U'_{g_1 g_2})^{-1} \text{,}
\end{equation}
where the transformed factor set now varies from site to site and is
\begin{equation}
\omega'_{\br}(g_1, g_2) = \lambda^{-1}_{\br}(g_1) \lambda^{-1}_{\br}(g_2) \lambda_{\br}(g_1 g_2) \omega(g_1, g_2) \text{.}
\end{equation}
This is a projective transformation of $\omega$, so $[\omega'_{\br}] = [\omega]$ and the fractionalization class is well-defined, despite the position-dependence of $\omega'_{\br}$.

\section{Gauging the symmetry}
\label{sec:gauging}

Gauging of internal unitary symmetry has emerged as a useful tool in the study of topological phases.\cite{levin12b}  One couples the system to a weakly-fluctuating $G$ gauge field, so if $G$ is discrete and $d \geq 2$, the resulting gauge theory is in a topologically ordered deconfined phase.  The  topological order after gauging can be used to distinguish different SPT phases, or different SET phases with the same topological order (before gauging).

We shall show that the result of gauging $G$ symmetry in our models is an $E$ gauge theory, where $E$ is the $\z_n$ central extension of $G$ for cohomology class $[\omega^e] \in H^2(G, \z_n)$  (defined below).  The $E$ gauge theory is that obtained by gauging a topologically trivial paramagnet with $E$ symmetry.  In addition, for the SET phases arising in the string flux models, we argue that isomorphism of central extensions (in an appropriate sense) corresponds to equivalence of SET phases.  This allows us to show that $[\omega^e]$ and $[\omega^e]^{-1}$ fractionalization classes give rise to the same phase (see also Sec.~\ref{sec:background}), but otherwise distinct fractionalization classes give rise to distinct SET phases.

Before proceeding to gauge the $G$ symmetry, we introduce the notion of central extension.  A $\z_n$ central extension of $G$ is a group $E$ satisfying $G = E / \z_n$, with $\z_n \subset E$ in the center of $E$.  An element $e \in E$ can be written $e = a u(g)$, where $a \in \z_n$, and $u(g)$ is a representative of $g \in G$ in $E$.  Taking the quotient by $\z_n$ identifies the coset $\{ a u(g) | a \in \z_n \}$ with $g \in G$.  The representative $u(g)$ is arbitrary up to $u(g) \to \lambda(g) u(g)$, where $\lambda(g) \in \z_n$, but once we make an arbitrary choice of representatives, pairs $a, u(g)$ give a unique labeling of all $e \in E$.  We have
\begin{equation}
u(g_1) u(g_2) = \omega(g_1, g_2) u(g_1 g_2) \text{,}
\end{equation}
where $\omega(g_1, g_2) \in \z_n$ is a $\z_n$ factor set, which satisfies the same associativity condition Eq.~(\ref{eqn:associativity}).  At this point the development entirely parallels the discussion of factor sets in Sec.~\ref{sec:preliminaries}, with projective transformations on $\omega(g_1, g_2)$ induced by $u(g) \to \lambda(g) u(g)$.

Now, to obtain the $E$ gauge theory by gauging $G$, we exploit the mapping to $\z_n$ gauge theory developed in Sec.~\ref{sec:gmapping}.
  We first observe that the $\z_n$ gauge theory can be obtained by starting with a trivial paramagnet, where we keep only the on-site degrees of freedom (\emph{i.e.} $b^\dagger_{\br g}$ and $\hat{N}_{\br g}$) from the gauge theory, and where the Hamiltonian is
\begin{equation}
H_{{\rm paramagnet}} =  - 2 \sum_{\br} \sum_{g \in G} \cos \Big( \frac{2\pi}{n} \hat{N}_{\br g} \Big)  \text{.}
\end{equation}
This paramagnet has $E$ symmetry, where $E$ is the $\z_n$ central extension of $G$ corresponding to $[\omega]$.  Writing $e = a u(g)$ for $e \in E$, the $E$ symmetry acts on $b^\dagger_{\br g_0}$ and $\hat{N}_{\br g_0}$by
\begin{eqnarray}
U^{\vphantom{-1}}_e b^\dagger_{\br g_0} U^{-1}_e &=& a\, \omega(g, g_0) b^\dagger_{\br, g g_0} \\
U^{\vphantom{-1}}_e \hat{N}_{\br g_0} U^{-1}_e &=& \hat{N}_{\br, g g_0} \text{.}
\end{eqnarray}
Upon gauging the subgroup $\z_n \subset E$, we obtain the $\z_n$ gauge theory $H_{{\rm gauge}}$ [Eq.~(\ref{eqn:Hgauge})].  Rather than first gauging $\z_n$ and then gauging $G$, we can equivalently gauge the $E$ symmetry all at once, to obtain an $E$ gauge theory.

Two such gauge theories with gauge groups $E$ and $E'$ have the same topological order if $E \simeq E'$, where $\simeq$ denotes group isomorphism.  However, because these gauge theories are obtained from SET phases with $\z_n$ topological order and $G$ symmetry, they have structure beyond just their topological order.  Indeed, it turns out that group isomorphism $E \simeq E'$ is not the right criterion to establish equivalence of the underlying (un-gauged) SET phases.  To proceed, we need to expose the additional structure of the $E$ gauge theory, which corresponds to structure of the central extension beyond just the group structure of $E$.  

The additional structure is captured nicely by using a slightly different definition of central extension.  We now 
define a central extension as the short exact sequence
\begin{equation}
\begin{CD}
0 @>>> \z_n @>i>> E @>\pi>> G @>>> 0 \text{.}
\end{CD}
\end{equation}
The arrows are group homomorphisms, $0$ is the trivial group, and $i(\z_n)$ lies in the center of $E$.  Exactness of the sequence means the kernel of each homomorphism is equal to the image of the preceding homomorphism.  This agrees with the definition of central extension above if $i$ is viewed as the inclusion  $i: \z_n \to \z_n \subset E$, and $\pi$ as the quotient map $\pi : E \to E / \z_n$.

In addition to the group structure of $E$, the maps $i$ and $\pi$ are also part of the structure of an extension, and have important physical interpretations.  The map $i$ identifies the $\z_n$ gauge group of the SET phase as a subgroup of $E$, while $\pi$ assigns a global symmetry operation $g \in G$ to each $e \in E$.  Any equivalence of SET phases must preserve the gauge group $i(\z_n)$, and must also preserve the assignment of global symmetry operations provided by $\pi$.  This is accomplished by defining two $\z_n$ central extensions of $G$ to be equivalent when there exists a commutative diagram
\begin{equation}
\begin{CD}
0 @>>> \z_n @>i>> E @>\pi>> G @>>> 0 \\
@.  @VV{\cal A}_{\z_n}V @VV{\rho}V  @VV1V @. \\
0 @>>> \z_n @>i'>> E' @>\pi'>> G @>>> 0
\end{CD} \text{.}
\label{eqn:ext-equiv}
\end{equation}
The rows of the diagram are short exact sequences corresponding to the central extensions, and the vertical arrows are group isomorphisms.  Therefore, $E \simeq E'$ is a necessary but not sufficient condition for two extensions to be equivalent.  Commutativity of the left square implies $\rho[ i(\z_n)] = i'(\z_n)$, the statement that the isomorphism preserves the $\z_n$ gauge group.  Moreover, commutativity of the right square implies $g = \pi[ i(a) u(g) ] = (\pi' \circ \rho)[i(a) u(g)]$, so the isomorphism preserves the assignment of $g \in G$ to each $e \in E$.  We note that ${\cal A}_{\z_n}$ can be any automorphism of $\z_n$ (there are only two), but only the identity map $1: G \to G$ preserves the assignment of global symmetry operations.

We would now like to understand how the factor sets $\omega$ and $\omega'$ of two equivalent central extensions are related.  For ${\cal A}_{\z_n} = 1$, it can be shown that the two extensions are equivalent if and only if $\omega$ and $\omega'$ are related by a projective transformation; that is, if and only if $[\omega] = [\omega']$.  On the other hand, suppose ${\cal A}_{\z_n} = {\cal A}_-$, where ${\cal A}_-(a) = a^{-1}$ is the non-trivial automorphism of $\z_n$.  (Note that ${\cal A}_- = 1$ for $n=2$.)  In this case, it can be shown the two extensions are equivalent if and only if
\begin{equation}
\omega(g_1, g_2) = \lambda(g_1) \lambda(g_2) \lambda(g_1 g_2)^{-1} \big(\omega'(g_1, g_2)\big)^{-1}  \text{,}
\end{equation}
or, equivalently, $[\omega] = [\omega']^{-1}$.

String flux models with different $e$-particle fractionalization classes are thus in different SET phases, except that $[\omega^e]$ and $[\omega^e]^{-1}$ fractionalization classes give rise to the same phase.  The latter statement also follows from the relabeling of anyons in Eqs.~(\ref{eqn:e-relabel}, \ref{eqn:m-relabel}), as discussed in Sec.~\ref{sec:preliminaries}.  Indeed, the automorphism ${\cal A}_-$ corresponds to taking the inverse of $\z_n$ charges and fluxes, and thus to the same relabeling of anyons.  We close this section by noting that, after accounting for relabeling of anyons, the results here confirm that different fractionalization classes give rise to distinct SET phases, as argued previously in Ref.~\onlinecite{essin13}.

\section{Discussion}
\label{sec:discussion}

The string flux models directly link the classification and mathematical structure of fractionalization classes with a mechanism for fractionalization in SET phases.  This feature suggests a number of open issues and directions for future work.  For instance, it will be valuable to look for more realistic models in which the string flux mechanism operates.  In this regard, it may be fruitful to consider space group symmetry, in part because this is a discrete symmetry common in realistic models, unlike the discrete internal symmetry we considered here.  Along these lines, following the discussion in Sec.~\ref{sec:intro},
it is likely possible to view projected parton wavefunctions with non-trivial projective symmetry groups\cite{wen02} as string flux ground states.  This could lead to a better microscopic understanding of such states,  perhaps inspiring new mean field theories or variational approaches.

Another idea that can probably be developed further is the magnetic route to low-energy gauge structure discussed in Sec.~\ref{sec:gmapping}.  Work in this direction could help broaden our understanding of the circumstances under which gauge structure plays an important role.

We remark that the string flux models can be generalized to include anti-unitary time reversal symmetry.  We have not pursued this in detail, but a cursory investigation\cite{mhunpub} suggests that when $G$ includes anti-unitary operations and we choose the symmetry to act trivially on $\z_n$ gauge charge, the generalization proceeds in a straightforward fashion \emph{provided} the factor set $\omega(g_1, g_2)$ is real.  For factor sets that cannot be made real by projective transformations, the na\"{\i}ve generalization appears to fail.

It is also interesting to consider the possibility of generalizing string flux models to include phenomena beyond symmetry fractionalization and thus describe a wider range of SET phases.  These models have a transparent link between classification and physical properties.  Therefore, if they can be generalized, it should enhance our understanding of SET phases, and may illuminate new physical phenomena.

\acknowledgments

I am happy to acknowledge inspiring discussions with Michael Levin and Xiao-Gang Wen, and am especially grateful for discussions and related collaborations with Andrew Essin and Hao Song.  This work was supported by the David and Lucile Packard Foundation, and in part by the National Science Foundation under Grant No. PHYS-1066293 and the hospitality of the Aspen Center for Physics.

\appendix

\begin{widetext}

\section{Form of $\Lambda_g(s_\ell)$ and linear action of symmetry}
\label{app:lambda}

We recall that $\Lambda_g(s_\ell) \in \z_n$ was introduced as a phase factor in the symmetry transformation of $a_\ell$ for the quantum double model on the Cayley graph of G [Eq.~(\ref{eqn:al-transf})].  In order for $G$ to be a symmetry of the Hamiltonian, we showed [Eq.~\ref{eqn:lambda-symmetry-requirement})] $\Lambda_g(s_\ell)$ must satisfy
\begin{equation}
\omega(g_1, g_2) = \Lambda_g(g_1) \Lambda_g(g_2) \Lambda^{-1}_g (g_1 g_2) \omega(g g_1 g^{-1} , g g_2 g^{-1} ) \text{.}
\end{equation}
Here, we find the solution for $\Lambda_g(s_\ell)$ given in Eq.~(\ref{eqn:lambda}).  We show that this solution is consistent with the requirement that symmetry act linearly on operators, and moreover that $U_g$ can be chosen so that $U_{g_1} U_{g_2} = U_{g_1 g_2}$.

To find $\Lambda_g(s_\ell)$, we consider a projective representation $\Gamma(g)$ with factor set $\omega(g_1, g_2)$, and we consider two ways of multiplying out the product $\Gamma(g) \Gamma(g_1) \Gamma(g^{-1}) \Gamma(g) \Gamma(g_2) \Gamma(g^{-1})$.  First,
\begin{eqnarray}
\Gamma(g) \Gamma(g_1) \Gamma(g^{-1}) \Gamma(g) \Gamma(g_2) \Gamma(g^{-1})
&=& \Big\{ \Big[ \Gamma(g) \big[ \Gamma(g_1) [ \Gamma(g^{-1}) \Gamma(g) ] \Gamma(g_2) \big]  \Big] \Gamma(g^{-1}) \Big\}  \nonumber \\
&=& \omega(g^{-1}, g) \omega(g_1, g_2) \omega(g, g_1 g_2) \omega(g g_1 g_2, g^{-1} ) \Gamma( g g_1 g_2 g^{-1} ) \text{.}
\end{eqnarray}
Second,
\begin{eqnarray}
\Gamma(g) \Gamma(g_1) \Gamma(g^{-1}) \Gamma(g) \Gamma(g_2) \Gamma(g^{-1})
&=&  \Big[ \big[ [ \Gamma(g) \Gamma(g_1) ] \Gamma(g^{-1}) \big] \big[  [\Gamma(g) \Gamma(g_2) \ \Gamma(g^{-1}) \big] \Big] \nonumber \\
&=& \omega(g, g_1) \omega(g g_1, g^{-1} ) \omega(g, g_2) \omega(g g_2, g^{-1} ) \omega(g g_1 g^{-1}, g g_2 g^{-1} )
\Gamma( g g_1 g_2 g^{-1} )  \text{.}
\end{eqnarray}
Setting these two expressions equal gives
\begin{eqnarray}
\omega(g_1, g_2) &=& \omega^{-1} (g^{-1}, g) [  \omega(g, g_1) \omega(g g_1, g^{-1} ) ] [ \omega(g, g_2) \omega(g g_2, g^{-1} ) ] [ \omega(g, g_1 g_2) \omega(g g_1 g_2, g^{-1} ) ]^{-1}  \omega(g g_1 g^{-1}, g g_2 g^{-1} )  \nonumber \\
&=& [  \omega^{-1} (g^{-1}, g) \omega(g, g_1) \omega(g g_1, g^{-1} ) ] [ \omega^{-1} (g^{-1}, g) \omega(g, g_2) \omega(g g_2, g^{-1} ) ] \nonumber \\
&\times& [ \omega^{-1} (g^{-1}, g) \omega(g, g_1 g_2) \omega(g g_1 g_2, g^{-1} ) ]^{-1}  \omega(g g_1 g^{-1}, g g_2 g^{-1} ) \text{.}
\end{eqnarray}

We have thus found
\begin{equation}
\Lambda_g(s_\ell) =  \omega^{-1} (g^{-1}, g) \omega(g, s_\ell) \omega(g s_\ell, g^{-1} ) \text{.} \label{eqn:lambda-3w}
\end{equation}
This expression can be simplified by noting that
\begin{equation}
\omega^{-1}(g^{-1}, g) = \omega(g, s_\ell) \omega(g^{-1}, g s_\ell) \text{,}
\end{equation}
which follows from associativity for the product $\Gamma(g^{-1}) \Gamma(g) \Gamma(s_\ell)$.  We then have the result given in Eq.~(\ref{eqn:lambda}):
\begin{equation}
\Lambda_g(s_\ell) = \omega^{-1}(g^{-1}, g s_\ell ) \omega(g s_\ell, g^{-1} ) \text{.}
\end{equation}

The requirement that symmetry act linearly on local operators [Eq.~(\ref{eqn:weak-linear-action})] leads to a condition on $\Lambda_g(s_\ell)$ given in Eq.~(\ref{eqn:lambda-condition}).  Before checking that this condition holds, we will first give an explicit form for $U_g$.  For this form of $U_g$, we will see that Eq.~(\ref{eqn:lambda-condition}) actually implies $U_{g_1} U_{g_2} = U_{g_1 g_2}$, which is a stronger statement than Eq.~(\ref{eqn:weak-linear-action}).

We first define $\tilde{U}_g$, which is a unitary operator that moves the link $\ell$ to $g \ell$, but has no action within the link Hilbert space.  Acting on operators,
\begin{equation}
\tilde{U}_g  a_\ell \tilde{U}^{-1}_g = a_{g \ell} \text{,} \quad \tilde{U}_g  e_\ell \tilde{U}^{-1}_g = e_{g \ell} \text{.}
\end{equation}
To completely specify ${U}_g$, we also need to consider its action on states.  We consider a product state where each edge $\ell$ is in the state $ | \alpha_\ell \rangle_\ell$, that is $| \psi \rangle = \prod_{\ell \in E} \otimes | \alpha_\ell \rangle_\ell$.  Then
\begin{equation}
\tilde{U}_g | \psi \rangle = \prod_{\ell \in E} \otimes | \alpha_\ell \rangle_{g \ell} = \prod_{\ell \in E} \otimes | \alpha_{g^{-1} \ell} \rangle_{\ell} \text{.}
\end{equation}
This just expresses precisely the statement that $\tilde{U}_g$ only moves edges around, with no action in the edge Hilbert space, and no phase factors.  It is clear that $\tilde{U}_{g_1} \tilde{U}_{g_2} = \tilde{U}_{g_1 g_2}$.

We let $k_g(s_\ell)$ be an integer defined modulo $n$, so that
\begin{equation}
\exp\Big( \frac{2 \pi i}{n} k_g(s_\ell) \Big) = \Lambda_g(s_\ell) \text{.}
\end{equation}
Then we define
\begin{equation}
U_g = \tilde{U}_g \prod_\ell (e^\dagger_{\ell})^{k_g(s_\ell)} \text{.}  \label{eqn:ug}
\end{equation}
It is easily checked that $U_g$ acts as desired on $e_\ell$ and $a_\ell$.  We now compute the product
\begin{eqnarray}
U_{g_1} U_{g_2} &=&  \tilde{U}_{g_1} [ \prod_\ell (e^\dagger_{\ell})^{k_{g_1} (s_\ell)} ] \tilde{U}_{g_2}  [ \prod_\ell (e^\dagger_{\ell})^{k_{g_2} (s_\ell)} ] \nonumber \\
&=& \tilde{U}_{g_1 g_2} \prod_{\ell} (e^\dagger_{\ell})^{ k_{g_1} (g_2 s_\ell g^{-1}_2) + k_{g_2} (s_\ell) } \text{.}
\end{eqnarray}
This equals $U_{g_1 g_2}$ provided that
\begin{equation}
k_{g_1} (g_2 s_\ell g^{-1}_2) + k_{g_2} (s_\ell) = k_{g_1 g_2} (s_{\ell}) \mod n \text{.}
\end{equation}
Using the definition of $k_g(s_\ell)$, this is equivalent to
\begin{equation}
\Lambda_{g_1} (g_2 s_\ell g_2^{-1} ) \Lambda_{g_2}(s_\ell) = \Lambda_{g_1 g_2} (s_\ell) \text{,} \label{eqn:lambda-condition-app}
\end{equation}
which is the same as Eq.~(\ref{eqn:lambda-condition}).

We now check this equation holds by obtaining it from an elementary identity involving commutators in group theory.  We observe that
\begin{equation}
\Lambda_{g^{-1}} ( g s_\ell ) = \omega(s_\ell, g) \omega^{-1}(g, s_\ell) \text{.}
\end{equation}
Recalling the commutator of two group elements $a, b$ is defined by $[a , b] = a b a^{-1} b^{-1}$, we have the identity
\begin{eqnarray}
\left[ \Gamma(s_\ell), \Gamma(g) \right] &=&  \omega(s_\ell, g) \omega^{-1}(g, s_\ell) \Gamma(s_\ell g) \Gamma^{-1}(g s_\ell) \nonumber \\
 &=& \Lambda_{g^{-1}}(g s_\ell) \Gamma(s_\ell g) \Gamma^{-1}(g s_\ell) \text{.}  \label{eqn:gamma-commutator}
\end{eqnarray}
In order to apply this result,   we put $g_1 \to g^{-1}_1$, $g_2 \to g^{-1}_2$, and $s_\ell \to g_2 g_1 s_\ell$ in Eq.~(\ref{eqn:lambda-condition-app}), obtaining the equivalent formula to be shown:
\begin{equation}
\Lambda_{g^{-1}_1} ( g_1 s_\ell g_2 ) \Lambda_{g^{-1}_2} ( g_2 g_1 s_\ell) = \Lambda_{g^{-1}_1 g^{-1}_2} ( g_2 g_1 s_\ell) \text{.}
\label{eqn:simpler-lambda-condition}
\end{equation}

For elements $a,b,c$ of any group, the following commutator identity holds:
\begin{equation}
 \left[ a b, c \right] \left[ c a , b \right] = \left[ a, b c \right]  \text{.}
\end{equation}
In particular,
\begin{equation}
\left[ \Gamma(s_\ell) \Gamma(g_2), \Gamma(g_1) \right] \left[ \Gamma(g_1) \Gamma(s_\ell), \Gamma(g_2) \right]
= \left[ \Gamma(s_\ell) , \Gamma(g_2) \Gamma(g_1) \right] \text{.}
\end{equation}
The products of two $\Gamma$'s can be combined, with the resulting factor sets canceling out, to obtain
\begin{equation}
\left[ \Gamma(s_\ell g_2), \Gamma(g_1) \right] \left[ \Gamma(g_1 s_\ell), \Gamma(g_2) \right]
= \left[ \Gamma(s_\ell) , \Gamma(g_2 g_1) \right] \text{.}
\end{equation}
Using Eq.~(\ref{eqn:gamma-commutator}) to evaluate the commutators, we have
\begin{equation}
\Lambda_{g^{-1}_1}(g_1 s_\ell g_2) \Lambda_{g^{-1}_2}(g_2 g_1 s_\ell) \Gamma(s_\ell g_2 g_1) \Gamma^{-1}(g_1 s_\ell g_2) \Gamma(g_1 s_\ell g_2) \Gamma^{-1}(g_2 g_1 s_\ell) = \Lambda_{g^{-1}_1 g^{-1}_2} (g_2 g_1 s_\ell) \Gamma(s_\ell g_2 g_1) \Gamma^{-1}(g_2 g_1 s_\ell) \text{.}
\end{equation}
The products of $\Gamma$'s on the two sides of the equations are equal, so we cancel out the $\Gamma$'s to obtain Eq.~(\ref{eqn:simpler-lambda-condition}).

\section{Ground states are invariant under $U_g$}
\label{app:gsug}

Here, we show that if $|\psi_{gs} \rangle$ is a ground state of one of the $d$-dimensional solvable models constructed in Sec.~\ref{sec:gend}, then, for all $g \in G$,
\begin{equation}
U_g | \psi_{gs} \rangle = | \psi_{gs} \rangle  \text{.} \label{eqn:gsinvt}
\end{equation}

We construct a ground state starting from
\begin{equation}
|\psi_0 \rangle = \prod_{\ell \in E} \otimes | a^0_\ell \rangle_{\ell} \text{,} \label{eqn:psi0}
\end{equation}
which is an eigenstate of $a_\ell$ for all $\ell \in E$ (with eigenvalue $a^0_\ell$), satisfying $B_p | \psi_0 \rangle = \omega_p | \psi_0 \rangle$.  Then, as discussed in Sec.~\ref{sec:anygraph}, we can obtain a ground state from $| \psi_0 \rangle$ by
\begin{equation}
\ket{\psi_{gs}} = \frac{1}{\sqrt{n}} \prod_{v \in V} \Big[ \frac{1}{\sqrt{n}} \sum_{a = 0}^{n-1} (A_v)^a \Big] \ket{\psi_0} \text{.} \label{eqn:onegs-app}
\end{equation}
The state $|\psi_0 \rangle$ can be varied, corresponding to threading $\z_n$ flux through the $d$ ``holes'' in the torus, and in this way one obtains a basis for the $n^d$-fold degenerate ground state subspace.  Verifying Eq.~(\ref{eqn:gsinvt}) for these basis states implies that it holds for any ground state.

We have
\begin{eqnarray}
U_g | \psi_{gs} \rangle &=& \frac{1}{\sqrt{n}} U_g  \prod_{v \in V} \Big[ \frac{1}{\sqrt{n}} \sum_{a = 0}^{n-1} (A_v)^a \Big] U^{-1}_g U_g \ket{\psi_0} \nonumber \\
&=& \frac{1}{\sqrt{n}}  \prod_{v \in V} \Big[ \frac{1}{\sqrt{n}} \sum_{a = 0}^{n-1} (A_{g v})^a \Big]  U_g \ket{\psi_0} \nonumber \\
&=& \frac{1}{\sqrt{n}}  \prod_{v \in V} \Big[ \frac{1}{\sqrt{n}} \sum_{a = 0}^{n-1} (A_{v})^a \Big]  U_g \ket{\psi_0} \text{.}  \label{eqn:uggs}
\end{eqnarray}
Now, from the form of $U_g$ in Eq.~(\ref{eqn:ug}), it is clear that $U_g | \psi_0 \rangle = | \psi'_0 \rangle$, where $| \psi'_0 \rangle$ is another state of the same form given in Eq.~(\ref{eqn:psi0}), only with different $a_\ell$-eigenvalues.  By symmetry, we still have $B_p | \psi'_0 \rangle = \omega_p | \psi'_0 \rangle$.  In fact, for an arbitrary cycle $c$ (including non-contractible cycles), $|\psi_0 \rangle$ and $|\psi'_0\rangle$ have the same eigenvalue of $B_c$.
All $\z_n$ fluxes in $|\psi_0\rangle$ and $|\psi'_0\rangle$ are thus the same, with only the $\z_n$ ``vector potential'' $a_\ell$ differing between these two states, so the two states are related by a gauge transformation.  This is expressed by writing
\begin{equation}
| \psi'_0 \rangle = \Big[ \prod_{v \in V} (A_v)^{G_v} \Big] | \psi_0 \rangle \text{,} \label{eqn:gt}
\end{equation}
where $G_v$ is an integer $0 \leq G_v \leq n-1$.

Starting from Eq.~(\ref{eqn:uggs}), Eq.~(\ref{eqn:gt}) implies
\begin{equation}
U_g | \psi_{gs} \rangle = \Big[ \prod_{v \in V} (A_v)^{G_v} \Big] | \psi_{gs} \rangle = | \psi_{gs} \rangle \text{,}
\end{equation}
where the last equality follows from the fact that $A_v | \psi_{gs} \rangle = | \psi_{gs} \rangle$ for all $v \in V$.

\section{Proof that $U^e_g$ gives a symmetry localization}
\label{app:eloc}

Here, we show that
\begin{equation}
U^e_g[ v(\br, g_0) ] =  \left\{  \begin{array}{ll} 
a_{(\br ; g_0, g g_0)}   & , \quad g \neq 1 \\
1  & , \quad g = 1 \end{array}\right. \text{,} \label{eqn:ueg-app}
\end{equation}
given in Eq.~(\ref{eqn:ueg}), gives a localization of $G$ symmetry in the solvable models.  That is, we need to show that
\begin{equation}
U_g | \psi_e \rangle = U^e_g(v_1) \cdots U^e_g(v_n) | \psi_e \rangle
\label{eqn:eloc-general-app}
\end{equation}
holds for any state $|\psi_e \rangle$ with $e$-particles at vertices $v_1, \dots, v_n$.  For $i = 1, \dots, n$, we write $v_i = v(\br_i, g_i)$.

\begin{figure}
\includegraphics[width=0.6\columnwidth]{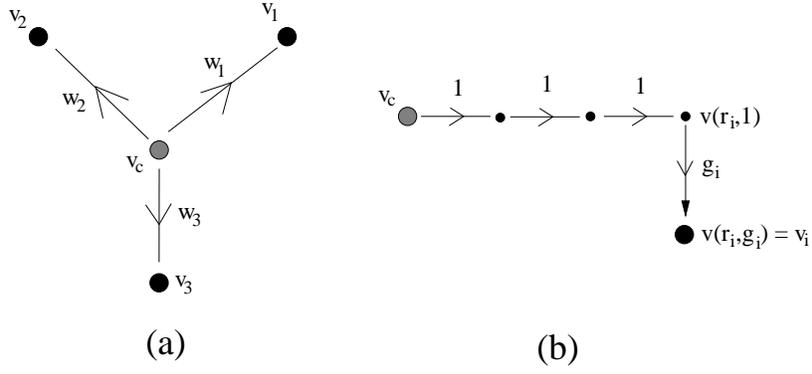}
\caption{(a) Depiction of $e$-strings used to construct a state $| \psi_e \rangle$, for $n=3$.  The $w_i$ string ($i = 1,2,3$) is oriented away from the central vertex $v_c$, toward each vertex $v_i$ where an $e$-particles resides.  (b)  Detail of the path $w_i$.  A path $w^s_i$ of spatial $1$-edges joins $v_c$ to $v(\br_i, 1)$, which is then joined to $v_i$ by the Cayley $g_i$-edge $(\br_i; 1, g_i)$.}
\label{fig:psie}
\end{figure}

First, we represent the state $| \psi_e \rangle$ by a product of $e$-string operators acting on a ground state.  We introduce a ``central'' vertex $v_c = v(\br_c, 1)$, and define paths $w_i$ joining $v_c$ to $v_i$ ($i = 1,\dots, n$), with orientation pointing away from $v_c$.  The path $w_i$ is the union of a path $w^s_i$ of spatial $1$-edges joining $v_c$ to $v(\br_i, 1)$, together with the Cayley edge $(\br_i; 1, g_i)$ that joins $v(\br_i, 1)$ to $v_i$ (see Fig.~\ref{fig:psie}).  Therefore
\begin{equation}
S^e(w_i) = S^e(w^s_i) a_{(\br_i ; 1, g_i)} \text{.}
\end{equation}

\begin{figure}
\includegraphics[width=0.35\columnwidth]{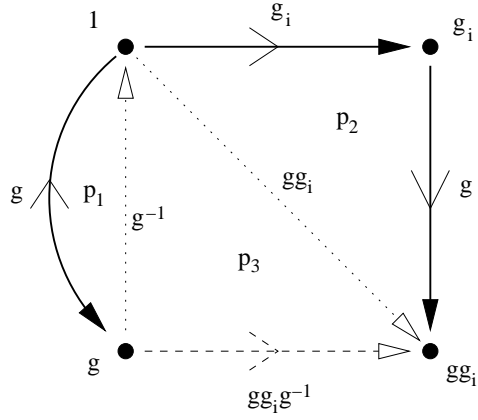}
\caption{Solid lines show the $e$-string $a^{-1}_{(\br_i, 1, g)} a_{(\br_i, 1, g_i)} a_{(\br_i, g_i, g g_i)}$, with site labels $\br_i$ suppressed.  This string is moved through the three plaquettes $p_1, p_2, p_3$, to coincide with the dashed-line string $a_{(\br_i; g, g g_i)}$ joining the same initial and final vertices.  The phase factors accumulated as the string passes through $p_1, p_2, p_3$ are $\omega^{-1}(g^{-1}, g)$, $\omega(g, g_i)$ and $\omega(g g_i, g^{-1})$, respectively.}
\label{fig:moving-string}
\end{figure}

The state $| \psi_e \rangle$ is given by
\begin{equation}
| \psi_e \rangle = \prod_{i = 1}^n S^e(w_i) | \psi_{gs} \rangle = \prod_{i = 1}^n [ S^e(w^s_i) a_{(\br_i ; 1, g_i) } ] | \psi_{gs} \rangle \text{,}
\end{equation}
where $|\psi_{gs} \rangle$ is a ground state.  Any state with $n$ $e$-particles at the specified vertices, and no other excitations, can be written as a linear combination of such states, where the terms in the linear combination differ only by different choices of the ground state $| \psi_{gs} \rangle$.

The left-hand side of Eq.~(\ref{eqn:eloc-general-app}) can be written
\begin{equation}
U_g | \psi_e \rangle =  \prod_{i = 1}^n [ S^e(g w^s_i) \Lambda_{g}(g_i) a_{(\br_i ; g, g g_i) } ] | \psi_{gs} \rangle \text{,} \label{eqn:eloc-lhs}
\end{equation}
where we used $U_g | \psi_{gs} \rangle = | \psi_{gs} \rangle$, as shown in Appendix~\ref{app:gsug}.  The right-hand side of Eq.~(\ref{eqn:eloc-general-app}) is
\begin{equation}
U^e_g(v_1) \cdots U^e_g(v_n) | \psi_e \rangle =  \big[ \prod_{i = 1}^n  S^e(w_i) a_{(\br_i, g_i, g g_i)}  \big] | \psi_{gs} \rangle  \text{.} \label{eqn:eloc-rhs}
\end{equation}
We proceed by bringing this expression to the form Eq.~(\ref{eqn:eloc-lhs}).

The $i$th term of the product in Eq.~(\ref{eqn:eloc-rhs}) is $S^e(w_i) a_{(\br_i, g_i, g g_i)} = S^e(w^s_i) a_{(\br_i, 1, g_i)} a_{(\br_i, g_i, g g_i)} $, which is an $e$-string operator with end points at $v_c$ and $v(\br_i, g g_i)$.  We thus have freedom to move the string while keeping its endpoints fixed, while keeping track of any phase factors accumulated.  We accomplish this in a few steps.  First, we have
\begin{equation}
S^e(w^s_i) a_{(\br_i, 1, g_i)} a_{(\br_i, g_i, g g_i)}  | \psi_{gs} \rangle
= a_{(\br_c, 1, g)} S^e(g w^s_i) a^{-1}_{(\br_i, 1, g)} a_{(\br_i, 1, g_i)} a_{(\br_i, g_i, g g_i)} | \psi_{gs} \rangle \text{.}
\end{equation}
This corresponds to moving each spatial $1$-edge in $w^s_i$ through a type II plaquette, to become a spatial $g$-edge.  This produces two additional Cayley edges, at the ends of $w^s_i$.  No phase factors are acquired, since $B_p = 1$ for type II plaquettes (acting on $| \psi_{gs} \rangle$).  Therefore,
\begin{eqnarray}
U^e_g(v_1) \cdots U^e_g(v_n) | \psi_e \rangle &=& \Big[ \prod_{i = 1}^n a_{(\br_c, 1, g)} S^e(g w^s_i) a^{-1}_{(\br_i, 1, g)} a_{(\br_i, 1, g_i)} a_{(\br_i, g_i, g g_i)} | \psi_{gs} \rangle \\
&=& \Big[ \prod_{i = 1}^n S^e(g w^s_i) a^{-1}_{(\br_i, 1, g)} a_{(\br_i, 1, g_i)} a_{(\br_i, g_i, g g_i)} | \psi_{gs} \rangle \text{,} \label{eqn:eloc-step1}
\end{eqnarray}
where the second equality holds since $(a_\ell)^n = 1$.

To finish bringing Eq.~(\ref{eqn:eloc-step1}) into the desired form Eq.~(\ref{eqn:eloc-lhs}), we need only deal with the $a_\ell$ factors contained within each Cayley subgraph at $\br_i$.  The operator $a^{-1}_{(\br_i, 1, g)} a_{(\br_i, 1, g_i)} a_{(\br_i, g_i, g g_i)}$ is an $e$-string with initial vertex $v(\br_i, g)$ and final vertex $v(\br_i, g g_i)$.  By moving the string through the three plaquettes shown in Fig.~\ref{fig:moving-string} while keeping the end points fixed, we have
\begin{equation}
a^{-1}_{(\br_i; 1, g)} a_{(\br_i; 1, g_i)} a_{(\br_i; g_i, g g_i)} | \psi_{gs} \rangle = \omega^{-1}(g^{-1}, g) \omega(g, g_i) \omega(g g_i, g^{-1} ) a_{(\br_i; g, g g_i) } | \psi_{gs} \rangle  = \Lambda_g(g_i) a_{(\br_i; g, g g_i) } | \psi_{gs} \rangle  \text{,}
\end{equation}
where the last equality follows from the expression for $\Lambda_g(s_\ell)$ in Eq.~(\ref{eqn:lambda-3w}).  Therefore,
\begin{equation}
U^e_g(v_1) \cdots U^e_g(v_n) | \psi_e \rangle = \Big[ \prod_{i = 1}^n S^e(g w^s_i)  \Lambda_g(g_i) a_{(\br_i; g, g g_i) }  | \psi_{gs} \rangle \text{,}
\end{equation}
and we have established the desired result.

\section{Consistency of $U_g$ in quantum double model and gauge theory}
\label{app:ug-gauge}

In Sec.~\ref{sec:gmapping}, we established a correspondence between a low-energy limit of the string flux models, and $\z_n$ gauge theories.  It needs to be shown that $U_g$ as defined in the gauge theory gives the same action on gauge-invariant operators as $U_g$ in the quantum double model.  While this is straightforward in other cases, it requires some algebra for $a_\ell$ with $\ell$ a Cayley edge, which we present here.

Consider the Cayley $s_\ell$-edge $\ell = (\br; g_0, s_\ell g_0)$, so
\begin{equation}
U_g a_{(\br; g_0, s_\ell g_0) } U_g^{-1} = \Lambda_g(s_\ell) a_{(\br; g g_0 , g s_\ell g_0 ) } \text{.} \label{eqn:qdm-transformation}
\end{equation}
According to the operator dictionary, in particular Eq.~(\ref{eqn:al-cayley-corr}), the corresponding operator in the gauge theory is given by
\begin{equation}
a_{(\br; g_0, s_\ell g_0) }  \leftrightarrow \omega(s_\ell, g_0) b^\dagger_{\br , s_\ell g_0} b^{\vphantom\dagger}_{\br g_0} \text{.}
\end{equation}
Using the gauge theory definition of $U_g$ [Eq.~(\ref{eqn:gauge-ug})], we have
\begin{eqnarray}
U_g [ \omega(s_\ell, g_0) b^\dagger_{\br , s_\ell g_0} b^{\vphantom\dagger}_{\br g_0} ] U^{-1}_g
&=& \omega(s_\ell, g_0) \omega(g, s_\ell g_0) \omega^{-1}(g, g_0) b^\dagger_{\br, g s_\ell g_0} b^{\vphantom\dagger}_{\br, g g_0} \\
&=& \Big[ \omega(s_\ell, g_0) \omega(g, s_\ell g_0) \omega^{-1}(g, g_0) \omega^{-1}(g s_\ell g^{-1}, g g_0) \Big] \Big[ \omega(g s_\ell g^{-1}, g g_0) b^\dagger_{\br, g s_\ell g_0} b^{\vphantom\dagger}_{\br, g g_0} \Big] \\
&\leftrightarrow& 
\Big[ \omega(s_\ell, g_0) \omega(g, s_\ell g_0) \omega^{-1}(g, g_0) \omega^{-1}(g s_\ell g^{-1}, g g_0) \Big] a_{(\br; g g_0, g s_\ell g_0)} \text{.}
\end{eqnarray}
In order for this to be consistent with Eq.~(\ref{eqn:qdm-transformation}), we must have the relation
\begin{equation}
\omega(s_\ell, g_0) \omega(g, s_\ell g_0) \omega^{-1}(g, g_0) \omega^{-1}(g s_\ell g^{-1}, g g_0) = \Lambda_g(s_\ell) \equiv   \omega^{-1}(g^{-1}, g s_\ell ) \omega(g s_\ell, g^{-1} )\text{,} \label{eqn:gt-transf-toshow}
\end{equation}
which we now establish.

We proceed by manipulating the left-hand side of Eq.~(\ref{eqn:gt-transf-toshow}), which we define to be $F(g, s_\ell, g_0)$.  We consider $\Gamma(g)$, a projective representation of $G$ with factor set $\omega$, as a convenient means to derive associativity relations for $\omega$.  Associativity of the product $\Gamma(g) \Gamma(s_\ell) \Gamma(g_0)$ implies
\begin{equation}
\omega(s_\ell, g_0) \omega(g, s_\ell g_0) = \omega(g, s_\ell) \omega(g s_\ell, g_0) \text{.} \label{eqn:gtt1}
\end{equation}
Associativity of $\Gamma(g s_\ell g^{-1}) \Gamma(g) \Gamma(g_0)$ implies
\begin{equation}
\omega(g, g_0) \omega(g s_\ell g^{-1}, g g_0) = \omega(g s_\ell g^{-1}, g) \omega( g s_\ell, g_0) \text{.} \label{eqn:gtt2}
\end{equation}
Using Eqs.~(\ref{eqn:gtt1}) and~(\ref{eqn:gtt2}), we have
\begin{eqnarray}
F(g, s_\ell, g_0) &=&
[ \omega(s_\ell, g_0) \omega(g, s_\ell g_0) ] [ \omega(g, g_0) \omega(g s_\ell g^{-1}, g g_0) ]^{-1} \\
&=& [  \omega(g, s_\ell) \omega(g s_\ell, g_0) ] [ \omega(g s_\ell g^{-1}, g) \omega( g s_\ell, g_0) ]^{-1} \\
&=&  \omega(g, s_\ell) \omega^{-1}(g s_\ell g^{-1}, g) \text{,} \label{eqn:gtt3}
\end{eqnarray}
and the apparent dependence on $g_0$ has disappeared.

Next, associativity of $\Gamma(g s_\ell) \Gamma(g^{-1}) \Gamma(g)$ implies
\begin{equation}
\omega(g s_\ell g^{-1}, g) = \omega^{-1}(g s_\ell, g^{-1}) \omega(g^{-1}, g) \text{,}
\end{equation}
and, therefore,
\begin{equation}
F(g, s_\ell, g_0) = \big[ \omega(g, s_\ell) \omega^{-1}(g^{-1}, g) \big] \omega(g s_\ell, g^{-1}) \text{.}
\end{equation}
Finally, associativity of $\Gamma(g^{-1}) \Gamma(g) \Gamma(s_\ell)$ implies
\begin{equation}
\omega(g, s_\ell) \omega^{-1}(g^{-1},g) = \omega^{-1}(g^{-1}, g s_\ell) \text{,}
\end{equation}
so
\begin{equation}
F(g, s_\ell, g_0) = \omega^{-1}(g^{-1}, g s_\ell)  \omega(g s_\ell, g^{-1}) = \Lambda_g(s_\ell) \text{,}
\end{equation}
the desired result.

\end{widetext}

\bibliography{cgrefs}

\end{document}